# Late-end reionization with ATON-HE: towards constraints from Ly $\alpha$ emitters observed with *JWST*

Shikhar Asthana ,[1]★ Martin G. Haehnelt,[1] Girish Kulkarni ,[2] Dominique Aubert,[3] James S. Bolton [4] and Laura C. Keating [5]

[1]*Kavli Institute for Cosmology and Institute of Astronomy, Madingley Road, CB3 0HA Cambridge, UK*
[2]*Tata Institute of Fundamental Research, Homi Bhabha Road, 400005 Mumbai, India*
[3]*Observatoire Astronomique de Strasbourg, 11 rue de l'Universite, F-000 Strasbourg, France*
[4]*School of Physics and Astronomy, University of Nottingham, University Park, NG7 2RD Nottingham, UK*
[5]*Institute for Astronomy, University of Edinburgh, Blackford Hill, EH9 3HJ Edinburgh, UK*

**ABSTRACT**
We present a new suite of late-end reionization simulations performed with ATON-HE, a revised version of the GPU-based radiative transfer code ATON that includes helium. The simulations are able to reproduce the Ly $\alpha$ flux distribution of the E-XQR-30 sample of QSO absorption spectra at $5 \lesssim z \lesssim 6.2$, and show that a large variety of reionization models are consistent with these data. We explore a range of variations in source models and in the early-stage evolution of reionization. Our fiducial reionization history has a mid-point of reionization at $z = 6.5$, but we also explore an 'Early' reionization history with a mid-point at $z = 7.5$ and an 'Extremely Early' reionization history with a mid-point at $z = 9.5$. Haloes massive enough to host observed Ly $\alpha$ emitters are highly biased. The fraction of such haloes embedded in ionized bubbles that are large enough to allow high Ly $\alpha$ transmission becomes close to unity much before the volume filling factor of ionized regions. For our fiducial reionization history this happens at $z = 8$, probably too late to be consistent with the detection by *JWST* of abundant Ly $\alpha$ emission out to $z = 11$. A reionization history in our 'Early' model or perhaps even our 'Extremely Early' model may be required, suggesting a Thomson scattering optical depth in tension with that reported by Planck, but consistent with recent suggestions of a significantly higher value.

**Key words:** radiative transfer – galaxies: high-redshift – intergalactic medium – quasars: absorption lines – (galaxies:) quasars: general – dark ages, reionization, first stars.

## 1 INTRODUCTION

The Epoch of Reionization (EoR) marks a critical juncture in the cosmic history when the intergalactic medium (IGM) reionizes due to ultraviolet radiation from the earliest cosmic sources (Furlanetto, Peng Oh & Briggs 2006; Dayal & Ferrara 2018). As a vital epoch in the chronology of the Universe, the EoR serves as a bridge, linking the smooth matter distribution of the early Universe to the large-scale structure observed in the present day (Loeb & Furlanetto 2013). Observational studies of the EoR paint a picture of a complex process. The onset of reionization occurs around UV sources, leading to the creation of ionized bubbles. Over time, these bubbles gradually coalesce, culminating in a completely ionized IGM, an idea first introduced in Arons & McCray (1970; see reviews by Ciardi & Ferrara 2005; Choudhury & Ferrara 2006 for more details). Moreover, additional studies underscore the inherent inhomogeneity of the reionization process, as exhibited in transmission spikes observed in the Ly $\alpha$ forest within quasar spectra at $z > 5$ (Miralda-Escude, Haehnelt & Rees 2000; Becker et al. 2015; Bosman et al. 2022).

Attempting to delineate this phase in time, several observational probes have been employed to constrain the onset and end of the EoR. Observations of Thomson scattering of CMB photons by free electrons formed during reionization offers valuable insights into the timing of this epoch (Kogut et al. 2003; Planck Collaboration VI 2020). The completion of reionization is now believed to have occurred between $z = 5$ and 6, inferred from the accelerated evolution of the Ly $\alpha$ optical depth at $z > 5$ (Becker et al. 2001; Fan et al. 2006), concomitant with the observed decline of Ly $\alpha$ emission from high-redshift galaxies (Stark et al. 2010; Choudhury et al. 2015).

To gain a thorough understanding of this epoch, various teams have performed a range of diverse simulations to replicate the later stages of reionization. These efforts can be grouped into two categories: the first involving coupled hydrodynamics and radiative transfer (Gnedin 2014; Rosdahl et al. 2018; Ocvirk et al. 2020; Kannan et al. 2021), and the second using the technique of post-processing hydrodynamical simulations with radiative transfer (Iliev et al. 2006; Aubert & Teyssier 2010; Eide et al. 2018; see review by Gnedin & Madau 2022). In post-processing simulations, the absence of coupling between radiation and hydrodynamics results in the omission of the hydrodynamic response of the IGM to inhomogeneous photoheating, and limits the detailed modelling of galaxy populations and the escape of ionizing radiation. None the less, as the pressure smoothing scale of the IGM is small, neglecting the radiation-hydrodynamic coupling does not significantly affect the spatial distribution of ionized and neutral regions on large scales (Kulkarni et al. 2015; Puchwein et al. 2023).

---

★ E-mail: sa2001@cam.ac.uk

On the other hand, these simulations offer computational advantages, being more economically feasible. They are particularly effective in reproducing the large-scale characteristics of the Ly $\alpha$ forest during and post-reionization, provided that the ionizing emissivity is well calibrated (Kulkarni et al. 2019a; Keating et al. 2020a). Note also that more recently Puchwein et al. (2023) presented a hybrid scheme where the result of hydrosimulations post-processed with ATON were used to capture the hydrodynamic response of the IGM to inhomogeneous reionization iteratively by re-running the hydrosimulations with inhomogeneous photoheating rates derived from the post-processed simulations, offering an interesting compromise.

The work by Kulkarni et al. (2019a) and Keating et al. (2020a) utilized the hydrogen-only radiative transfer code ATON, with a single photon frequency, to construct models that accurately reproduce the substantial spatial fluctuations in the effective optical depth of the Ly $\alpha$ absorption in QSO spectra at $z > 5$ (Bosman et al. 2018). These simulations, however, did not include helium, and the ad hoc increase in photon energy invoked to imitate the effect of helium, along with the lack of multiple frequencies, somewhat restricted the scope of these simulations. More precise comparisons between the IGM temperature at $z < 6$ and observations will necessitate a more comprehensive model of the IGM. Such a model should encapsulate both hydrogen and helium, along with their interactions with higher energy photons (Ciardi et al. 2012). More recent data also asks for better modelling. Bosman et al. (2022) have presented measurements of the effective Ly $\alpha$ opacity of the IGM at redshifts 5–6 from high resolution, high SNR quasar spectra from the E-XQR-30 data set (D'Odorico et al. 2023). This data set is the combination of quasar spectra found by the ESO Large Programme 'XQR-30', and 12 spectra from the XSHOOTER archive. These new measurements offer more stringent constraints on the distribution of neutral islands in the early Universe and extends the observational reach to higher redshifts. Consequently, these enhancements in observational data quality necessitate a re-evaluation and recalibration of existing models to ensure that they reproduce the most current and accurate observations.

Here, we build on the work of Kulkarni et al. (2019a) and Keating et al. (2020a) by incorporating helium into ATON and conduct multifrequency simulations calibrated to the mean Ly $\alpha$ flux measurements by Bosman et al. (2022). We also investigate the effect of the rarity of the sources, studying a range of source populations and their contributions to the overall photon budget necessary for reionization. It is widely assumed that galaxies serve as the primary sources of ionizing photons (Bunker et al. 2004; Bouwens et al. 2015; Finkelstein et al. 2019; Ocvirk et al. 2021), whereas high-redshift active galactic nuclei are assumed to have a limited impact (Shapiro & Giroux 1987; Songaila, Cowie & Lilly 1990; Meiksin & Madau 1993; Ciardi et al. 2000; Bolton et al. 2005; Volonteri & Gnedin 2009; Fontanot, Cristiani & Vanzella 2012; Madau & Haardt 2015; Qin et al. 2017; Kulkarni, Worseck & Hennawi 2019b). However, uncertainties about the specific galaxy types involved and their associated mass ranges remain. Following Cain et al. (2023), in addition to our fiducial model, we developed simulations employing 'Oligarchic' and 'Democratic' source models. The Oligarchic model evaluates reionization driven by rarer bright sources, and the Democratic model assesses the possibility that the total ionizing emissivity is dominated by lower mass faint galaxies (see Hassan et al. 2022 for a recent discussion of the effect of different source models on reionization).

*JWST* has opened a new chapter in reionization studies with its much improved spectroscopic capabilities allowing to obtain high-quality spectra of reionization epoch galaxies and active galactic nuclei (AGNs) to $z = 11$ and beyond. Galaxies emitting Ly $\alpha$ radiation are an important probe of the neutral hydrogen fraction of the IGM (Zheng et al. 2017; Mason et al. 2018, 2019; Nakane et al. 2024; Umeda et al. 2024). Recent spectroscopic observations conducted by *JWST* have confirmed the presence of galaxies emitting Ly $\alpha$ radiation at a redshift of $z \sim 11$ (Tacchella et al. 2023). Ly $\alpha$ emitting galaxies must inhabit large ionized regions with radii $\gtrsim 0.5$ pMpc (Miralda-Escudé 1998; Mason & Gronke 2020; Hayes & Scarlata 2023) to avoid scattering of the Ly $\alpha$ photons out of the line of sight by the parts of the IGM where hydrogen is still neutral (Mesinger et al. 2014; Weinberger et al. 2018; Keating et al. 2024; Tang et al. 2024; Witstok et al. 2024). To study the prospects of constraining earlier stages of reionization, we construct three reionization models, with a deliberately wide range of early stage evolution. In addition to our fiducial simulation that has a mid-point of reionization at $z \sim 6.5$, we construct an 'Early' model with a midpoint of reionization at $z \sim 7.5$, a new 'Extremely Early' model with a mid-point of reionization at $z \sim 9.5$. All three models are calibrated to the Ly $\alpha$ mean transmission at redshift $\lesssim 6$. The fiducial model is similar to the model of Kulkarni et al. (2019a) and the fiducial model of Keating et al. (2020a). We will see below that only the 'Early' and fiducial models are consistent with the Thomson scattering optical depth reported by Planck (Planck Collaboration VI 2020), but see recent work by de Belsunce et al. (2021) and Giarè, Di Valentino & Melchiorri (2023) for suggestions that the Thomson scattering optical depth may need to be revised upwards.

The paper is structured as follows. Section 2 describes our simulation set-up. Section 3 discusses our calibration approach and examines the effects of the number of frequency bins, inclusion of helium, and source SED on various observables. Section 4 explores variations of the source model. In Section 5, we assess the impact of varying the mid-point of reionization. Section 6 describes our method for calculating ionized bubble sizes, discusses the source distributions within ionized regions, and draws implications for the visibility of LAE at high redshifts with *JWST*. The broader context of our findings is discussed in Section 7, comparing our work with other studies in the field. The paper concludes in Section 8, summarizing our main findings. Throughout this work, we assume a Lambda cold dark matter cosmology with parameter values from Planck Collaboration XVI (2014) ($\Omega_m = 0.308$, $\Omega_\lambda = 0.6982$, $h = 0.678$, $\Omega_b = 0.482$, $\sigma_8 = 0.829$, and $n = 0.961$).

## 2 SIMULATION SET-UP

Our reionization simulations are performed by post-processing cosmological hydrodynamical simulations for radiative transfer using ATON-HE, which is a modified version of the ATON code. ATON-HE incorporates helium into ATON and overhauls the multifrequency implementation of the original code.

### 2.1 Cosmological hydrodynamical simulations

We post-process simulations from the Sherwood–Relics suite of simulations (Puchwein et al. 2023) that were run using the Tree-PM SPH code P-GADGET-3. The suite consists of several different box sizes, out of which we use the box with size 160 cMpc/$h$ for our models. The simulations have $2 \times 2048^3$ gas and dark matter particles. In our previous work, we have identified 160 cMpc/$h$ as the 'sweet spot' for calibration of $2048^3$ simulations to the flux PDF (Kulkarni et al. 2019a; Keating et al. 2020b). Note, however, that as discussed in more detail in Feron et al. (2024) even at this dynamic range simulations with a box size of 160 cMpc/$h$ are not fully converged with regard to either resolution or box size. The simulations are initialized at a redshift $z = 99$, and snapshots are saved at 40 Myr intervals down to redshift $z = 4$. Star formation in these simulations is reduced to a simplified recipe (invoked using the QUICK_LYALPHA compile-time flag in P-GADGET-3): when gas particles exceed a density threshold of $\Delta = 10^3$, with a temperature $\lesssim 10^5$ K, they are removed from the hydrodynamic calculations and converted into star particles (Viel, Haehnelt & Springel 2004). As ATON-HE requires a uniform Cartesian grid, we project the gas density on to a grid of $2048^3$ cells.

## 2.2 Radiative transfer with ATON

ATON is a radiative transfer code that reduces the dimensionality of the radiative transfer equation by taking angular moments of the cosmological radiative transfer equation (Aubert & Teyssier 2008, 2010). The equations are truncated at second-order using the M1 relation (Levermore 1984), creating a closed system of equations. This approach converts the radiative transfer equations into a set of equations for energy and flux. In this paper, we ran ATON with multiple frequency bins and incorporated helium. We refer to this revised version of ATON as ATON-HE, and provide a detailed description of this code in Appendix A. In an ATON-HE simulation, after establishing the initial gas and source distribution, sources are placed at the locations of dark matter haloes following a source model. The spectrum of each source is modelled using a black-body spectrum. The temperature of the black-body spectrum allocated to the ionizing sources, the breakdown of this spectrum into discrete frequency bins, and the associated total volume emissivity, are input parameters for ATON-HE.

### 2.2.1 Source distribution

In order to model the ionizing sources we assign ionizing emissivities to the haloes identified in the simulations. For this we extract two properties of the haloes from the simulations: the mass and the number of haloes in each mass bin. The mass function of the haloes and its evolution with redshift is illustrated in Fig. 1, with colour representing the redshift. We see that the simulation struggles to resolve haloes with mass below $\sim 10^9$ M$_\odot$ $h^{-1}$. We therefore employ a cut-off of $10^9$ M$_\odot$ $h^{-1}$ on halo mass and only use haloes more massive than this in our source modelling. The vertical line in Fig. 1 shows this halo mass cut-off. In our fiducial simulations, the first set of sources greater than the cut-off mass form at $z = 19.55$, with a minimum and maximum halo mass of $10^9$ M$_\odot$ $h^{-1}$ and $1.8 \times 10^9$ M$_\odot$ $h^{-1}$. As the simulation proceeds, by $z = 6$, the minimum mass remains the same while the maximum halo mass increases to $4.1 \times 10^{13}$ M$_\odot$ $h^{-1}$.

### 2.2.2 Source spectrum

Fig. 2 illustrates the black-body source spectrum, its discretization, and the effect of the assumed black-body temperature used to characterize the ionizing spectrum of sources in our simulations. The figure depicts black-body spectra for two distinct temperatures $4.0 \times 10^4$ and $7.0 \times 10^4$ K, each discretized for computation into energy bins, and highlighted by different shaded regions. For hydrogen-only simulations, using a single-frequency bin is often sufficient as e.g. discussed by Keating, Puchwein & Haehnelt (2018) and by Kulkarni et al. (2019a). However, for helium, which has three ionization states, a single frequency approach is not adequate. The spectrum is thus divided based on the ionization energies of hydrogen and helium, represented by the solid vertical curves at 13.6, 24.6, and 54.4 eV in Fig. 2. We conducted a convergence test, as detailed further in Appendix D, and found that with four energy bins the thermal evolution is sufficiently converged when modelling both hydrogen and helium (cf. Mirocha et al. 2012).

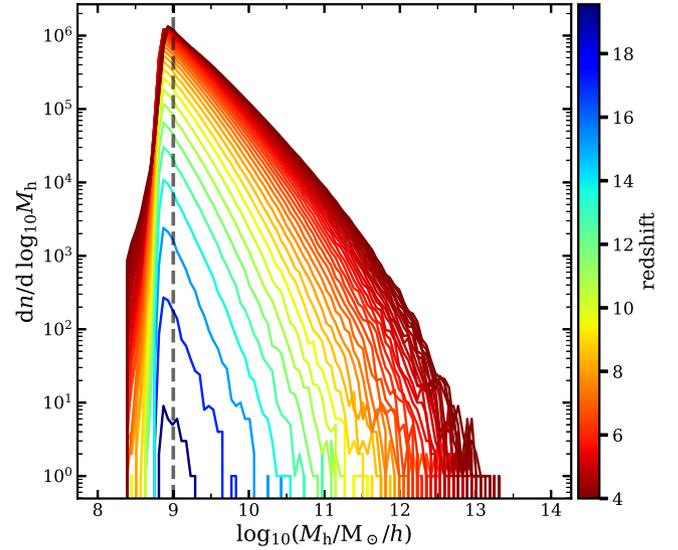

**Figure 1.** The total halo mass function in our simulation. The dashed line shows the total halo mass cut-off at $10^9$ M$_\odot$ $h^{-1}$ that we impose in our fiducial source model, as the simulation struggles to resolve haloes with mass below this value.

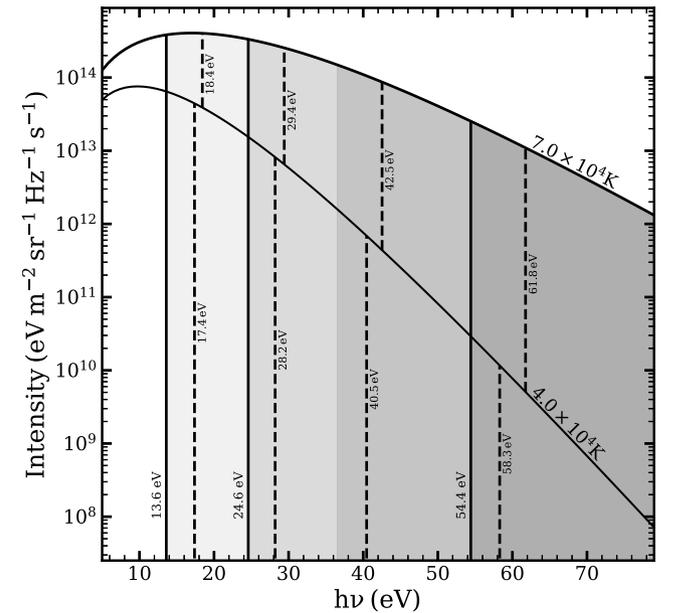

**Figure 2.** The black-body spectrum for two values of temperature. These spectra are used as source SEDs in our simulations. Shaded regions indicate our chosen frequency bins, each represented by a distinct colour. The highest energy bin extends to infinity. Dotted curves within each shaded region represent the average energy value for the corresponding bin. We have also marked the ionization energies for H I, He I, and He II.

After discretizing the spectrum, the next task is to ascertain the average energy for the photons within each bin, given by

$$\bar{E} = E_i + E_{\text{excess}}$$
$$= E_i + \frac{\int_{E_i}^{E_{i+1}} dE_\nu (I_\nu/E_\nu)(E_\nu - E_i)}{\int_{E_i}^{E_{i+1}} dE_\nu (I_\nu/E_\nu)}, \quad (1)$$

where $E_i$ is the lower end of an energy bin, and $I_\nu$ and $E_\nu$ are the specific intensity and energy at frequency $\nu$ (Pawlik & Schaye 2011). The dotted vertical lines in Fig. 2 represent this average energy for each energy bin. The average energy, as calculated by equation (1), is 17.4, 28.2, 40.5, and 58.3 eV for the black-body temperature of $4.0 \times 10^4$ K, and is 18.4, 29.4, 42.5, and 61.8 eV for the black-body temperature of $7.0 \times 10^4$ K.

### 2.2.3 Ionizing emissivity

An ionizing emissivity is assigned to each source halo. In our fiducial model, we follow Kulkarni et al. (2019a) and Keating et al. (2020b), and assume that the ionizing emissivity is proportional to the total halo mass. This is motivated by the fact that the star formation rate at these redshifts is expected to be approximately proportional to the halo mass. We set a total ionizing volume emissivity value, and distribute this emissivity over the haloes accordingly. We then calibrate the emissivity to obtain a range of evolutionary histories that are consistent with the observed mean Ly $\alpha$ transmission at $5 \lesssim z \lesssim 6.2$, as described in Section 3.1. A full simulation run from $z = 19.3$ to $z = 4.9$ for a multifrequency simulation including helium requires approximately 900 GPU-hours, utilizing 32 NVIDIA A100 GPUs.

### 2.3 The simulation suite

With the method described above, we created a suite of simulations, each varying in certain physical parameters, as detailed in Table 1. All simulations were conducted with a box size of 160 cMpc $h^{-1}$.

The first aspect we examined was the impact of the number of frequency bins, ranging from one to four. This led to simulations being categorized as either monofrequency or multifrequency. Additionally, we included simulations with and without helium to assess its influence on our results.

Another parameter of interest was the temperature of the black-body spectrum of the ionizing sources. We tested two different temperatures for this purpose: $4 \times 10^4$ and $7 \times 10^4$ K. This allowed us to study the effects of varying source spectra on the simulations.

We further investigated the role of the minimum mass cut-off and the mass distribution for sources emitting ionizing photons. In this context, we focused on three distinct models: fiducial, 'Democratic', and 'Oligarchic' (Cain et al. 2023). The fiducial model assumes that the emissivity of the sources is proportional to the total halo mass, with the minimum mass set at $10^9$ M$_\odot$ $h^{-1}$. The Democratic model, in contrast, assigns the same ionizing emissivity to all sources, regardless of the halo mass, thereby enabling us to assess the effect of small mass haloes. The Oligarchic model selects a higher mass cut-off and maintains a distribution proportional to the total halo mass, thereby allowing us to examine the role of high mass galaxies during reionization.

To complete our series of cosmological simulations, we vary the redshift of the midpoint of reionization, making it larger than in our fiducial model in an 'Early' and 'Extremely Early' model.

**Table 1.** Summary of simulations considered in this paper. The simulation box size is 160 cMpc $h^{-1}$, with $2048^3$ grid for the radiative transfer, corresponding to a spatial resolution of 78.125 ckpc $h^{-1}$. The spectra of sources in simulations were modelled to be black-body with temperatures of $4.0 \times 10^4$ K, and $7.0 \times 10^4$ K. The minimum mass of the haloes capable of producing ionizing photons is represented by $M_{\text{min}}$. The mid-point of reionization is represented by $z_{\text{mid}}$. The scaling of the ionizing photon emissivity $\dot{N}_{\text{ion}}$ of the sources with halo mass $M$ is also given.

| Simulation name | Source black-body temperature ($10^4$ K) | Number of frequency bins | IGM composition | Features |
|---|---|---|---|---|
| | | | | Simulations to quantify the effect of helium and multifrequency radiative transfer |
| Fiducial (Helium-40) | 4 | 4 | Hydrogen + Helium | $M_{\text{min}} = 10^9$ M$_\odot$ $h^{-1}$, $\dot{N}_{\text{ion}} \propto M$ |
| Mono-H–40 | 4 | 1 | Hydrogen | $M_{\text{min}} = 10^9$ M$_\odot$ $h^{-1}$, $\dot{N}_{\text{ion}} \propto M$ |
| Multi-H–40 | 4 | 4 | Hydrogen | $M_{\text{min}} = 10^9$ M$_\odot$ $h^{-1}$, $\dot{N}_{\text{ion}} \propto M$ |
| Mono-H–70 | 7 | 1 | Hydrogen | $M_{\text{min}} = 10^9$ M$_\odot$ $h^{-1}$, $\dot{N}_{\text{ion}} \propto M$ |
| Multi-H–70 | 7 | 4 | Hydrogen | $M_{\text{min}} = 10^9$ M$_\odot$ $h^{-1}$, $\dot{N}_{\text{ion}} \propto M$ |
| Helium-70 | 7 | 4 | Hydrogen + Helium | $M_{\text{min}} = 10^9$ M$_\odot$ $h^{-1}$, $\dot{N}_{\text{ion}} \propto M$ |
| | | | | Simulations with different source models |
| Oligarchic | 4 | 4 | Hydrogen + Helium | $M_{\text{min}} = 8.5 \times 10^9$ M$_\odot$ $h^{-1}$, $\dot{N}_{\text{ion}} \propto M$ |
| Democratic | 4 | 4 | Hydrogen + Helium | $M_{\text{min}} = 10^9$ M$_\odot$ $h^{-1}$, $\dot{N}_{\text{ion}}$ independent of halo mass |
| | | | | Simulations with different reionization histories |
| Early | 4 | 4 | Helium + Hydrogen | $M_{\text{min}} = 10^9$ M$_\odot$ $h^{-1}$, $\dot{N}_{\text{ion}} \propto M$, $z_{\text{mid}} = 7.5$ |
| Extremely early | 4 | 1 | Hydrogen | $M_{\text{min}} = 10^9$ M$_\odot$ $h^{-1}$, $\dot{N}_{\text{ion}} \propto M$, $z_{\text{mid}} = 9.5$ |

## 3 EFFECT OF HELIUM AND MULTIFREQUENCY RT

Using our simulations, we can now investigate the impact of helium and a multifrequency treatment of radiative transfer. Our corresponding results are shown in Fig. 3. By comparing with results from the Mono-H-40 and Mono-H-70 runs, the Multi-H-40 and Multi-H-70 runs allow us to isolate the effect of increasing the number of photon energy bins. Comparing with the Helium-40 and Helium-70 runs allows us to understand the role of helium. We focus on these six simulations in this section.

### 3.1 Calibration to the mean Ly $\alpha$ flux

We randomly extract 6400 distinct sight-lines from each simulation snapshot, at the same spatial resolution as that of the density grids used by ATON-HE. For these sight-lines, we calculate both the mean flux and the distribution of the effective optical depth along these sight-lines. Ly $\alpha$ optical depths were computed using the approximation to the Voigt profile as described in Tepper-García (2006). The mean flux in the simulated spectra is then calibrated to that from observed QSO absorption spectra at $z > 5.0$ as measured by Bosman et al. (2022). The volume emissivity is modified manually until the mean flux agrees with the observations for redshifts $5 \lesssim z \lesssim 6.2$. In Fig. 3(a), we show the evolution of the mean Ly $\alpha$ flux with redshift. The observed mean flux is shown as black points. We succeeded to reach good agreement with the observed data for all of our models. As we show in Appendix C, a 5–10 per cent disagreement in the mean flux between different calibrated runs is permissible. The effective optical depth $\tau_{\rm eff}$ is calculated by the conventional definition, $\langle F \rangle = e^{-\tau_{\rm eff}}$, measured in 50 cMpc $h^{-1}$ segments of the Ly $\alpha$ forest.

### 3.2 Ionizing volume emissivity

In Fig. 3(c), we show the redshift evolution of the volume emissivity of our models. A consistent feature across all models is the necessity of a drop in the emissivity at $z < 7$, which tends to level off as reionization approaches its conclusion, roughly at $z = 5$. Within the context of current models, this decline in emissivity remains challenging to explain (Cain et al. 2024), but see Ocvirk et al. (2021) and Qin et al. (2021) for discussions of this drop in the context of radiative suppression of star formation and modelling with 21CMFAST, respectively.

Note that in the six models that we discuss in this section we kept the evolution of the ionizing volume emissivity the same at $z \gtrsim 9$. Below $z = 9$, the multifrequency simulations fitting the observed Ly $\alpha$ forest data have a somewhat lower emissivity compared to their monofrequency counterparts. This lower emissivity is due to the combined effects of less absorption near the host halo due to the smaller cross-section for harder photons and the somewhat higher temperature and thus lower recombination rates in the multifrequency simulations.

Comparing our fiducial 'Helium-40' and the 'Helium-70' simulation with the multifrequency, hydrogen-only simulations 'Multi-H-40', and 'Multi-H-70', it is evident that the simulations including helium require a larger emissivity. The reason is the absorption of photons by helium, which otherwise would have ionized hydrogen. For this reason, at redshifts when the ionizing volume emissivity is still the same (at $z \gtrsim 9$), the hydrogen neutral fraction is the largest in the simulations including helium. To match the Ly $\alpha$ forest data at lower redshift, a larger emissivity is therefore required, which is why the simulations including helium require the highest emissivity at $z \lesssim 9$.

Fig. 3(d) shows the evolution of the ratio of the integrated number of ionizing photons per hydrogen atom as a function of redshift. Similar to the ionizing emissivity we find that the multifrequency simulations require the lowest number of photons per hydrogen atom ($\sim 2.3$), while the simulations including helium require the largest ($\sim 2.5$).

### 3.3 IGM temperature

Moving from the monofrequency, 'Mono-H-40' and 'Mono-H-70', to the corresponding multifrequency hydrogen-only simulations, 'Multi-H-40' and 'Multi-H-70', leads to a discernibly higher IGM temperature at mean density (Fig. 3f). This higher temperature for the same assumed source SED is due to the inclusion of higher energy photons in the multifrequency simulations that inject more energy into the IGM.

Conversely, the temperatures in the simulations in the fiducial 'Helium-40' and in the 'Helium-70' simulation are slightly lower than in their hydrogen-only multifrequency counterparts. This is due to the larger ionization energy of helium that results in a smaller amount of energy injected into the IGM, leading to lower temperatures, especially noticeable at $z > 9$. As already discussed up to $z \sim 9$ the models have the same ionizing volume emissivity.

The differences between the blue and green curves in the figure illustrate the effect of altering the source spectrum. Harder spectra with larger black-body temperature have a larger proportion of higher energy photons. The higher energy photons inject more energy into the IGM, explaining why the green curves consistently exceed the blue ones.

The peaks in the temperature-redshift curves thereby correlate closely with the reionization history. Notably, in the multifrequency hydrogen simulation 'Multi-H-70' with a black-body temperature of $7 \times 10^4$ K reionization finishes a little bit earlier than in the corresponding 'Helium-70' simulation, leading to the temperature curve peaking at slightly higher redshift. Note that an earlier conclusion of reionization leads to adiabatic cooling exceeding photoheating earlier. In contrast, with a source black-body temperature of $4 \times 10^4$ K, in the simulations with and without helium ('Multi-H-40' and the fiducial 'Helium-40', respectively) reionization completes at around the same time. Consequently, their temperature curves peak around the same redshift, but there is still a noticeable temperature difference as expected due to the presence or absence of helium. In essence, the exact timing of the peak in the temperature is governed by the conclusion of reionization, while the amplitude is influenced by the presence or absence of helium.

Fig. 3(f) shows a variety of temperature measurements as collected in Gaikwad et al. (2020), which served as a reference for selecting the temperature of the black-body spectrum assumed in our simulations. We aimed to encompass the range of measured temperatures, leading to the choice of black-body temperatures of $4 \times 10^4$ and $7 \times 10^4$ K. Note, however, that the temperatures in our simulations are higher than the measurements by Garzilli et al. (2017) and Walther et al. (2019), suggesting that there is still some uncertainty in what should be assumed for the source spectrum.

### 3.4 Neutral fraction and mean free path evolution

In Fig. 3(b), we show the evolution of the volume averaged neutral hydrogen. We also show observations from a variety of different probes focusing on estimates from dark gaps from Ly $\alpha$ forest data (McGreer et al. 2014; Zhu et al. 2022), the fraction of Lyman break galaxies showing Ly α emission (Mason et al. 2018, 2019), the damping wing analysis of QSOs proximity zones (Greig et al. 2017; Bañados et al. 2018; Davies et al. 2018; Greig et al. 2019; Wang et al. 2020; Yang et al. 2020a; Ďurovčíková et al. 2024), the analysis of Lyman-series opacities (Yang et al. 2020b; Gaikwad et al. 2023), the evolution of Ly α equivalent widths (Nakane et al. 2024), and the damping wing analysis from galaxies (Umeda et al. 2024). Our models are well within the error bars of these observations. Comparing our models we see that at $z \gtrsim 9$, the fraction of neutral hydrogen in the simulations including helium is consistently higher compared to that in the hydrogen-only simulations. As we assumed the same number of ionizing photons emitted in these models at high redshift this is not surprising and can be attributed to some higher energy photons, which were previously ionizing hydrogen now ionizing helium instead. Interestingly, the exact timing when reionization is completed varies between the different models despite all of them being calibrated to the observed mean Ly $\alpha$ flux at $5 \lesssim z \lesssim 6.2$ to well within the observational errors. As a consequence of these small variations in timing, there can be a difference of two orders of magnitude or more in the average neutral fraction when the IGM transitions rapidly to being highly ionized at $z = 5.5$ (compare e.g. the dashed green curve and the solid blue curves). We discuss these differences further in Section 3.5.

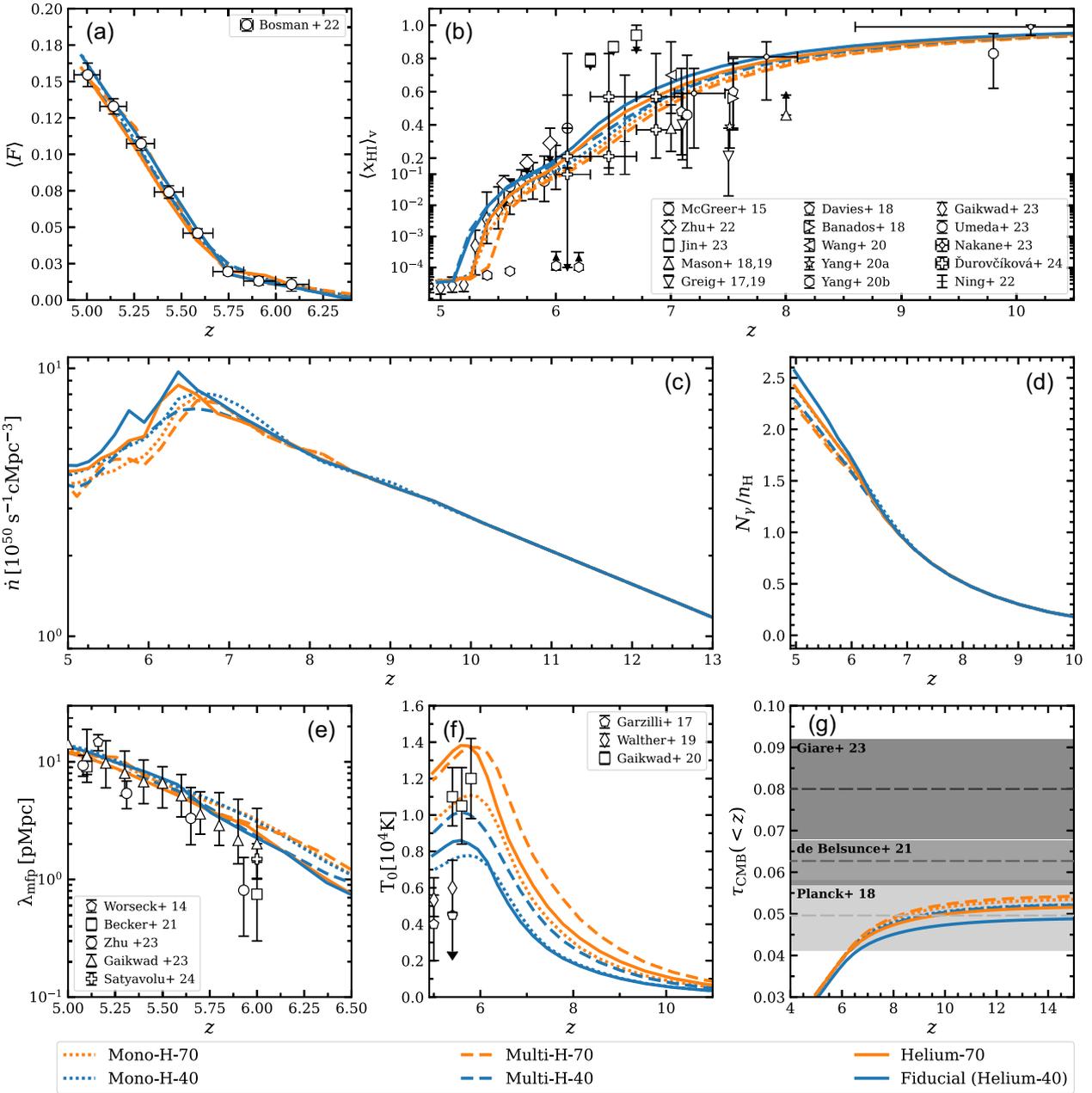

**Figure 3.** The reionization histories and various other related quantities from six simulations presented in this paper. Panel (a) shows the mean Ly α transmission compared with measurements by Bosman et al. (2022). Panel (b) shows the volume-averaged neutral hydrogen fraction $\langle x_{\rm HI}\rangle_{\rm v}$. This panel also shows inferences from dark gaps in the Ly α forest (McGreer, Mesinger & D'Odorico 2014; Zhu et al. 2022; Jin et al. 2023), the fraction of Lyman-break galaxies showing Ly α emission (Mason et al. 2018, 2019), quasar damping wings (Greig et al. 2017; Bañados et al. 2018; Davies et al. 2018; Greig, Mesinger & Bañados 2019; Wang et al. 2020; Yang et al. 2020a; Ďurovčíková et al. 2024), galaxy damping wings (Umeda et al. 2024), Ly α emission equivalent widths (Nakane et al. 2024), and the effective Ly α opacity of the IGM (Yang et al. 2020b; Ning et al. 2022; Gaikwad et al. 2023). Panel (c) shows the hydrogen-ionizing emissivity in the simulations. Panel (d) shows the ratio of the integrated number of hydrogen-ionizing photons to the number of hydrogen atoms. Panel (e) compares the mean free path of hydrogen-ionizing photons in the simulations with measurements by Worseck et al. (2014), Becker et al. (2021), Zhu et al. (2023), Gaikwad et al. (2023), and Satyavolu et al. (2023). Panel (f) plots the evolution of the IGM temperature at mean density in relation to measurements by Garzilli, Boyarsky & Ruchayskiy (2017), Walther et al. (2019) and Gaikwad et al. (2020). Panel (g) shows the cumulative Thomson scattering optical depth compared to measurements of Planck Collaboration VI (2020), de Belsunce et al. (2021), and Giarè et al. (2023) of the total Thomson scattering optical depth to the last scattering surface.

We have also calculated the average mean free path of Lyman-limit photons for all our models using the method discussed in Kulkarni et al. (2016). We take the average Lyman-continuum transmission across random sight-lines in the comoving frame, and fit it with an exponential with an e-folding length scale of $\lambda_{\rm mfp}$,

$$\langle \exp(-\tau_{912}) \rangle = F_0 \exp\left(-\frac{x}{\lambda_{\rm mfp}}\right), \qquad (2)$$

where $\lambda_{\rm mfp}$ is the mean free path, and $x$ is the position along a sight-line (Rybicki & Lightman 1986). The results are plotted in Fig. 3(e). The differences between the simulations are moderate and can be traced back to the differences in neutral fraction and the differences in spectral distribution of the ionizing photons which lead to differences in absorption cross-section. The mean free path in the simulations agree well with those measured by Gaikwad et al. (2023), but are about a factor two higher than those measured in the near-zones of QSOs at $5.8 < z < 6$ by Becker et al. (2021), Satyavolu et al. (2023), and Zhu et al. (2023). At lower redshift the observed measurements converge, and also agree well with our simulations. As expected the volume-averaged neutral fraction and the mean free path are anticorrelated. Looking more closely at models with similar reionization histories, e.g. the 'Mono-H-40' and the 'Helium-70' models, we see that the mean free path is shorter in the case of the simulation including helium.

### 3.5 Effective optical depth distribution

As discussed above, we have adjusted the ionizing volume emissivity so that the mean flux in the simulated spectra reproduces the mean observed Ly $\alpha$ flux. However, the full flux distribution of the effective optical depth contains more information. In Fig. 4, we show the cumulative distribution function (CDF) of the effective optical depth measured over 50 cMpc $h^{-1}$ segments for all ten distinct simulations spanning two redshift ranges: 5.36–5.51 and 5.67–5.83. For each simulation, the shaded regions denote the 15.87–84.13 percentiles of 10 000 realizations of the CDF, each created using the same number of LOS as the observed data, i.e. 65 and 48, respectively. The black curves show the observational data from Bosman et al. (2022). Notably, this newer data set offers tighter constraints on the effective optical depth CDF than the measurements by Bosman et al. (2018) and Eilers, Davies & Hennawi (2018; see in Fig. 4(l) for a comparison) with a less pronounced tail of high effective optical depth in the $5.67 < z < 5.83$ redshift bin, suggesting a somewhat smaller volume filling factor of neutral gas than the other two measurements at this redshift.

Overall, the agreement between observed and simulated flux CDFs is good, but matching the mean flux does not always translate into reproducing the width of the observed distribution, and in particular the tail at high optical depth. For both black-body temperatures the tail at high effective optical depth in the redshift range $5.67 \lesssim z \lesssim 5.83$ becomes more pronounced, as we move from monofrequency to hydrogen-only multifrequency simulation and multifrequency simulation including helium. Note however that in this redshift range, all six models nevertheless show comparable average neutral fractions.

In the redshift range between 5.36 and 5.51 shown as the left set of curves in each panel, all of our models fall somewhat short at high effective optical depths. The discrepancies observed among the models at this redshift can be attributed to variations in their average neutral fractions. The 'Multi-H-70' simulation, in particular, shows a significantly lower neutral fraction, which is reflected in its CDF extending to lower effective optical depth.

When examining the CDF of the effective optical depth, it is apparent that even if the mean flux values agree well across simulations, significant variations are present in their effective optical depth distributions. These variations are most noticeable in the tail end of the distribution curves. Due to the exponential relationship between mean flux and effective optical depth, higher values of effective optical depth contribute less significantly to the mean flux. This permits a wide range of variation in the high effective optical depth tail without markedly affecting the mean flux. Another intriguing observation is how small the differences in mean flux and effective optical depth distribution are, even when there is a substantial difference (up to two orders in magnitude or more) in the average neutral fraction between $5 \lesssim z \lesssim 5.5$. In this redshift range the average neutral fraction is typically quite low. In such cases, a limited number of neutral cells can significantly alter the average neutral fraction. Hence, a relatively small proportion of neutral cells strongly influences the overall average neutral fraction. Conversely, the mean flux along any given line of sight (LOS) is most significantly affected by the regions with the lowest density. Therefore, despite substantial variations in the average neutral fraction along a LOS, the mean flux may remain relatively constant, provided the least dense regions are similar. To summarize, the average neutral fraction is predominantly set by cells with the highest neutral fractions and hydrogen number densities. On the other hand, the flux in most of the volume is influenced more by the least dense cells. This dichotomy leads to the differences in the effective optical depth distribution and the volume averaged neutral fraction in simulations with similar mean flux.

Note that when calibrating simulations to the same mean flux, we also encounter a notable variation in the distribution of the effective optical depth CDF even for similar median effective optical depth. This is possible because once the effective optical depth in a 50 cMpc $h^{-1}$ segment of the spectrum is large the exact value has little effect on the median effective optical depth and the mean flux as the whole segment of spectrum consists mainly of saturated absorption.

### 4 VARYING THE SOURCE MODEL

In this section, we examine three distinct multifrequency helium models with a spectral black-body temperature of $4 \times 10^4$ K: fiducial, and following Cain et al. (2023), a 'Democratic' and a 'Oligarchic' source model. Similarly to Fig. 3, the properties of simulations with these different source models are compared to our fiducial model in Fig. 5 by the solid blue, dashed green and dashed purple curves. The fiducial 'Helium-40' model assigns an ionizing photon emissivity to haloes in proportion to their mass ($\dot{N}_{\rm ion} \propto M$). In the Democratic model, the total ionizing emissivity is evenly distributed across all haloes, regardless of their mass. This model allows us to assess the impact of smaller haloes in the reionization process. The Oligarchic model, on the other hand, biases the emissivity towards larger halo masses by assuming a higher mass cut-off for ionizing sources at $8.5 \times 10^9 \, {\rm M}_\odot \, h^{-1}$ instead of $10^9 \, {\rm M}_\odot \, h^{-1}$, thereby focusing on the influence of more massive galaxies in driving reionization. The emissivity distribution is similar to the fiducial simulations, i.e. $\dot{N}_{\rm ion} \propto M$.

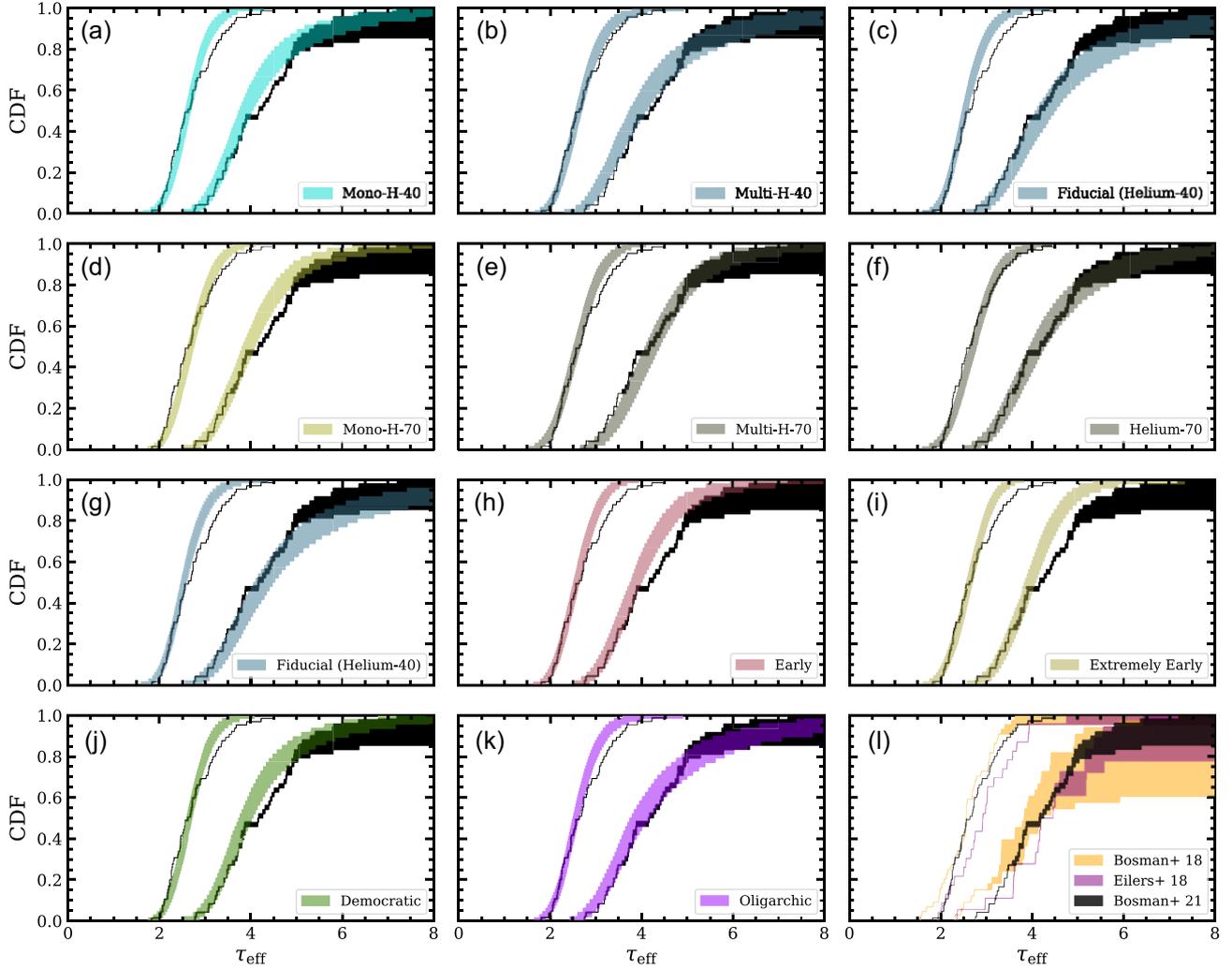

**Figure 4.** Panels (a–k) show the CDFs of the effective Ly α opacity $\tau_{\rm eff}$ of the IGM computed over 50 cMpc/$h$ sections of the Ly α forest in the ten simulations presented in this paper. Each of these ten panels shows two CDFs. The CDF with smaller average values of $\tau_{\rm eff}$ corresponds to $5.36 < z < 5.51$, and the other CDF corresponds to $5.67 < z < 5.83$. These shaded regions denote the 15.87–84.13 percentiles of 10 000 realizations of the CDF, each created using the same number of LOS as in the observed sample, i.e. 65 and 48, respectively, measured over 50 cMpc $h^{-1}$ segments of the Ly α forest. The black shaded regions in each of the panels show measurements by Bosman et al. (2022). Panel (l) compares measurements by Bosman et al. (2018), Eilers et al. (2018), and Bosman et al. (2022).

Comparing the fiducial 'Helium-40' model with the Oligarchic model we see that their emissivity (Fig. 5c) is similar at higher redshifts ($z \gtrsim 7$), while the Oligarchic model requires slightly less emissivity at lower redshifts to fit the Ly α forest data. Despite similar emissivities at $z \gtrsim 7$, the Oligarchic model exhibits a roughly 10 per cent lower neutral hydrogen fraction (panel b) around $z \sim 7$. This difference arises because the Oligarchic model with its higher mass cut-off assigns larger mass haloes an increased emissivity. The resultant earlier ionization also allows for more adiabatic cooling, slightly lowering the IGM temperature (panel f) at later times in the Oligarchic model compared to the fiducial model. The Oligarchic model also predicts a larger mean free path (panel e), suggesting a smaller number of remaining small neutral islands within the extensive ionized regions. Fig. 4(k) shows that the Oligarchic model is very similar to the fiducial model in terms of the distribution of effective depths across the observed redshift range.

Comparing the Democratic model with the fiducial 'Helium-40' model, we see that the emissivity (panel c) is again similar at $z \gtrsim 7$ by construction, while the Democratic model requires higher emissivity at lower redshifts to agree with the Ly α forest data. The reionization histories (panel b) of both models agree closely at $z \gtrsim 7$, but the increased emissivity below this redshift causes reionization to happen slightly faster in the Democratic model, resulting in a reduced neutral fraction. The slightly faster ionization in the Democratic model also manifests as a slightly earlier peak in the temperature profile (panel f) compared to the fiducial model. The mean free paths (panel e) for both models are almost identical, probably reflecting a similar number of remaining neutral islands within ionized zones as in the fiducial model. Fig. 4(j) illustrates that the CDF of the Democratic

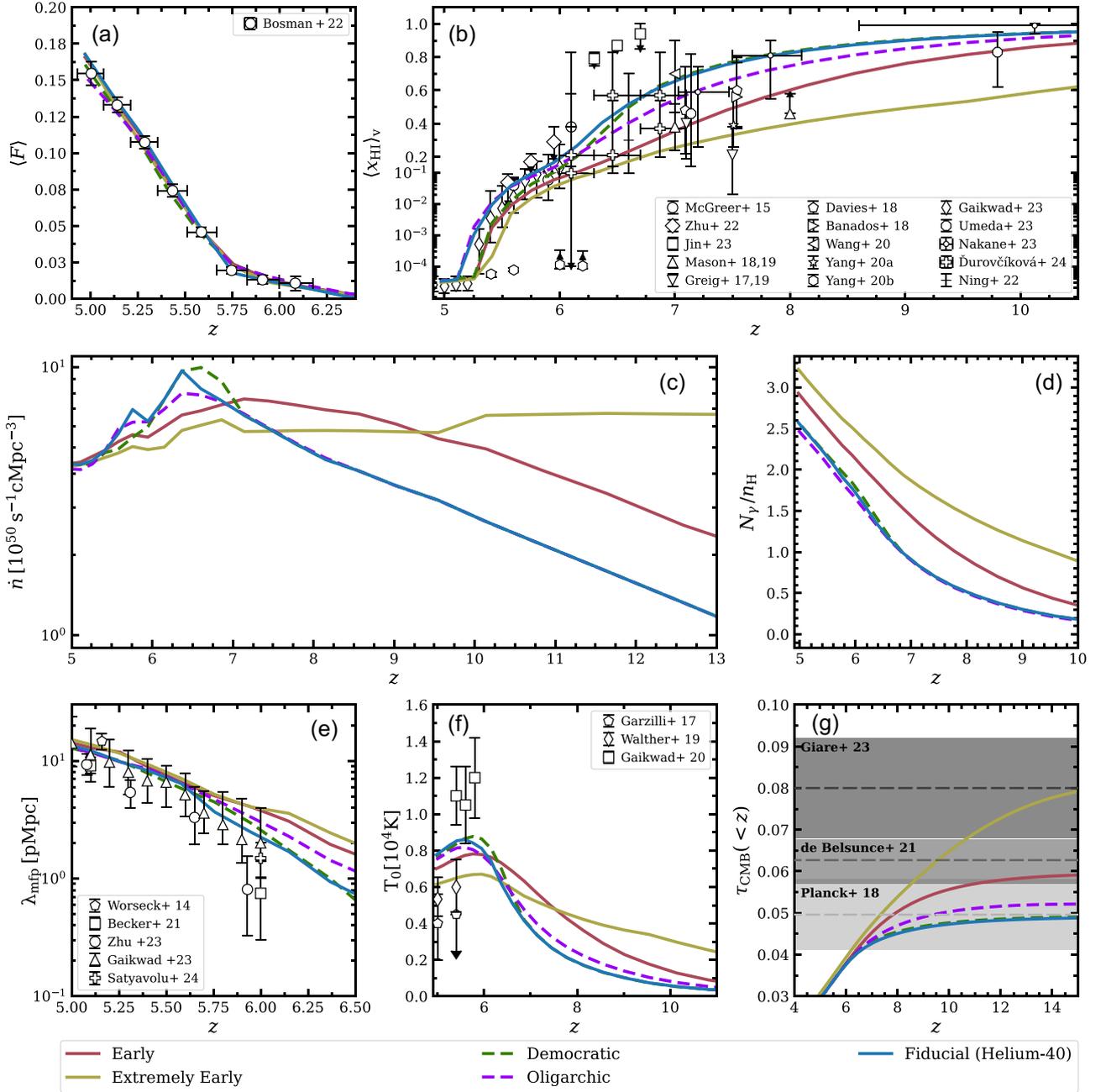

**Figure 5.** The reionization histories and various other related quantities as in Fig. 3 for the Democratic, Oligarchic, 'Early' and 'Extremely Early' reionization models. The observational points are the same as in Fig. 3.

model is similar to the fiducial model in both redshift ranges plotted.

Fig. 6 presents a two-dimensional slice depicting the neutral fraction for the three different source models at $z = 6.15$, a point where the volume-averaged neutral fractions of the models are similar: $\langle x_{HI}\rangle_v = 0.734, 0.798, 0.788$ for fiducial, Democratic, and Oligarchic models, respectively. Individually identifiable ionized regions in the Oligarchic model are notably larger, and have a lower neutral fraction, a reflection of the higher emissivity allocated to massive haloes. Conversely, the Democratic model exhibits more numerous smaller ionized bubbles. The differences in bubble morphology reflects the more spatially dispersed distribution of lower mass haloes and the more highly biased clustering of the higher mass haloes. The fiducial model falls between the other two source models, with the number of small ionized regions exceeding that in the Oligarchic model, but being lower than in the Democratic model.

In conclusion, we find that calibration to the mean flux is possible for very different source populations as long as the ionizing emissivity is properly chosen. In the Oligarchic model the required ionizing emissivity was thereby somewhat lower and in the Democratic model somewhat higher than for our fiducial source model.

## 5 THE EARLY STAGES OF REIONIZATION

While a late end of reionization is now well established thanks to the Ly $\alpha$ forest data, the earlier evolution of the neutral fraction is still rather uncertain. However, with *JWST* it has become possible to obtain high-quality spectra of galaxies and AGN in the reionization epoch all the way to $z = 11$ and beyond (Curtis-Lake et al. 2023; Tacchella et al. 2023; Tang et al. 2023; Umeda et al. 2024). The effect of the damping wings of the remaining neutral hydrogen IGM, either directly observed as damping wings or due to the resulting suppression of the Ly $\alpha$ emission from reionization-epoch galaxies and AGN, offers the exciting prospect to significantly improve these constraints as the number of *JWST* spectra increases and the modelling improves (Keating et al. 2024; Umeda et al. 2024). To aid these efforts, we consider, in addition to our fiducial reionization history with a mid-point of reionization at $z = 6.5$, two other models with mid-points at $z = 7.5$ and $z = 9.5$, respectively. These models are also shown in Fig. 5 and are the fiducial (dark blue curve), 'Early' (red curve), and 'Extremely Early' (dark yellow curve), respectively. The properties of all models can be found in Table 1. In our analysis, we utilized simulations all featuring a black-body source spectrum with the same spectral temperature of $4 \times 10^4$ K as in our fiducial model. The 'Early' model was run like the fiducial model as a multifrequency simulation including helium, whereas the 'Extremely Early' model was run as a monofrequency hydrogen-only simulation. As outlined in Section 3, the inclusion of helium introduces some moderate variations in the observables, but for the purpose of examining the impact of significantly earlier reionization timing that we are pursuing here, the monofrequency hydrogen simulation we performed should be sufficient. We thus decided not to invest the significant resources required to rerun and re-calibrate this model including helium.

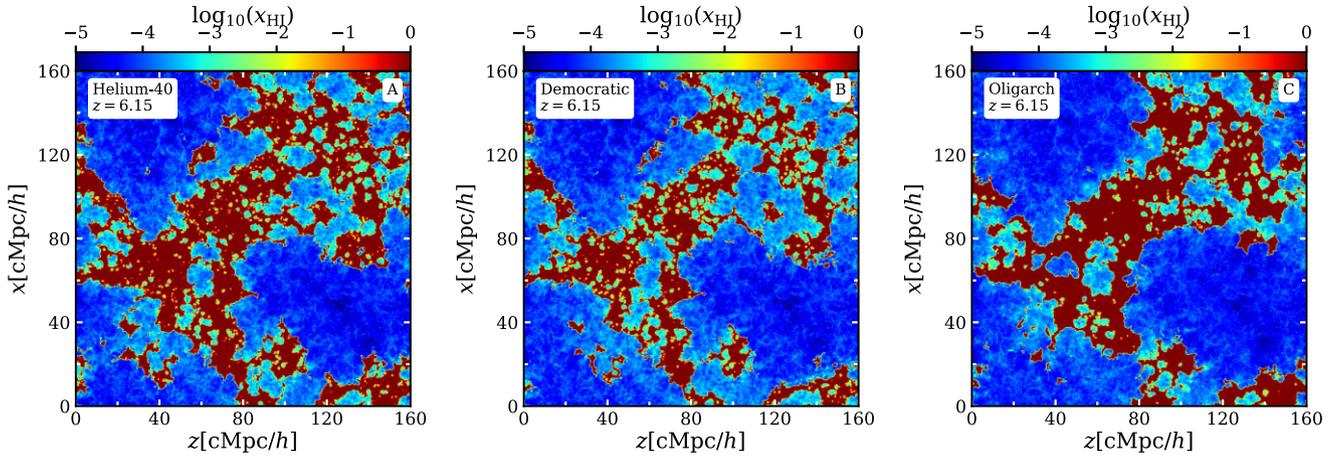

**Figure 6.** The distribution of neutral hydrogen fraction $x_{HI}$, shown in a two-dimensional slice from the neutral hydrogen distribution in the simulation box, in our fiducial ('Helium-40') simulation (left panel) and the simulations with the 'Democratic' and 'Oligarchic' source models (central and right panels).

The various properties of the 'Early' and 'Extremely Early' model are also shown in Fig. 5. In panel (a) the mean Ly $\alpha$ flux for the three reionization histories shows very small differences. Turning our attention to the emissivity, as shown in panel (c) of Fig. 5, we observe that the emissivity peaks at $z \sim 6.5$ for the fiducial, $z \sim 7.25$ for the 'Early', and $z \sim 10$ for the 'Extremely Early' model. The earlier peaks in the 'Early' and 'Extremely Early' simulation were required to start reionization earlier. The larger ionizing emissivity is reflected in a lower neutral fraction at high redshifts [panel (b) of Fig. 5]. In the 'Extremely Early' model by $z \sim 13$, the volume averaged neutral fraction is $\sim 0.8$, while the fiducial model still has a value of 0.99. As reionization proceeds, the neutral fraction of the three models start to decrease at different rates and by $z \sim 6$ it is 0.04 for the 'Extremely Early' model, while its 0.16 for the fiducial model. The differences persist down to $z = 5.2$, after which all three models have the same volume averaged neutral fraction. Comparing to the observational points, we see that the 'Extremely Early' model, at $z > 8$, might be more ionized than some of the data suggests. The effect of the difference in reionization timing is also seen for the temperature at mean density, shown in panel (f). As the 'Extremely Early' simulation reionizes earlier, it adiabatically cools for a longer duration leading to a lower value of the peak of the temperature at $z \sim 6$. The temperature peak of the 'Early' model is somewhat higher, while the temperature in the fiducial model is the highest, as it ionizes the fastest leaving less time for adiabatic cooling. When analysing the mean free path, it becomes clear that later reionization results in a somewhat shorter mean free path. Through the construction of these three models, we have effectively demonstrated that fitting to the Ly $\alpha$ mean flux still leaves room for a large variation in the reionization history. A greater number of models than perhaps initially anticipated are consistent with the mean Ly $\alpha$ flux data. Note, however, that the Thomson optical depth for the scattering of CMB photons for the 'Extremely Early' model is $> 3\sigma$ larger than reported by Planck Collaboration VI (2020), and is also significantly larger than those reported by de Belsunce et al. (2021), but is consistent with those recently reported by Giarè et al. (2023, see panel g).

In Fig. 7, we provide a visual representation of the fiducial 'Helium-40', and the 'Early' and 'Extremely Early' models, showing lightcones that chart the fraction of neutral hydrogen, starting at a redshift of 19 and extending beyond the completion of hydrogen reionization to redshift 5, as shown on the bottom *x*-axis. The sources in our models start producing ionizing photons as early as $z = 19$, and thus we start our lightcones from that particular redshift. Even though our simulation only encompasses a spatial dimension of $160 \, \text{cMpc} \, h^{-1}$, we have made use of its periodic boundary conditions to produce lightcones that exceed the dimensions of the simulation box, as shown on the *y*-axis. By orienting the lightcone at an angle through the simulation, we ensured varied viewing of the box but because of the rather large coherence length of ionized/neutral regions some periodicity remains. On the top axis, we show the corresponding age of the universe. Redder (bluer) colours in the lightcones correspond to more neutral (ionized) regions.

In the 'Extremely Early' model, the already substantial imposed emissivity at the highest redshifts combined with the scarcity of sources leads to some probably unrealistically luminous sources concentrated in a small number of regions that give rise to a few, but sizable, ionization bubbles at these redshifts. In the rightmost section of Fig. 7, these manifest themselves as blue regions around $z = 17.3$ in the 'Extremely Early' model (panel c), which slowly recombine as the simulation proceeds and our imposed emissivity is distributed over a larger number of ionizing sources. A closer examination of the bottom lightcone shows large ionized bubbles appearing as early as $z \sim 14$. However, these bubbles are still transient, primarily because of the elevated recombination rates characteristic of these redshifts, coupled with the redistribution of ionizing emissivity as larger numbers of sources become operational. This is a numerical artefact created due to our method of assigning ionizing emissivity to sources. The model settles quickly into a more realistic source population from $z \sim 12$ onwards, with larger number of ionizing sources sharing the imposed emissivity. Advancing through the simulation timeline, expansive ionized hydrogen zones take shape and merge, eventually resulting in a fully reionized universe by $z \sim 6$. This progression is reflected in the evolution of the volume-averaged neutral fraction shown in the bottom left of Fig. 5. In contrast, the fiducial model shown in the top lightcone of Fig. 7 is dominated by much smaller ionized regions (spanning only a few cMpc $h^{-1}$) that persist up to $z \sim 7.3$. Subsequently, there is a swift escalation in the expansion of these bubbles, which then quickly fill the entirety of the simulated volume. Notably, the fiducial reionization model retains extended pockets of dense, neutral hydrogen, enduring until $z \sim 5.5$. Positioned between these two extremes, the 'Early' model presents a more gradual growth of the ionized regions. The first ionized bubbles appear around $z \sim 9$, with significant neutral regions enduring until about $z \sim 6.4$. This disparity in neutral islands manifests as a notable difference in the CDF at high optical depth, as seen in the bottom four panels in Fig. 8. Here, we show the effective optical depth on the *x*-axis, and the CDF of the effective optical depth on the *y*-axis for eight redshift bins. The redshift range of each bin is shown at the bottom right and is the same as the redshift range of the observational points for the mean Ly α flux shown in Fig. 5(a). Mirroring our analysis in Section 3.5, we have extracted a sample of LOS from our snapshots that matched the observed number of sight-line for the respective redshift bins. This process was repeated ten thousand times, enabling us to estimate the $1\sigma$ (68.26 percent) range for the effective optical depth CDF for the different redshift bins.

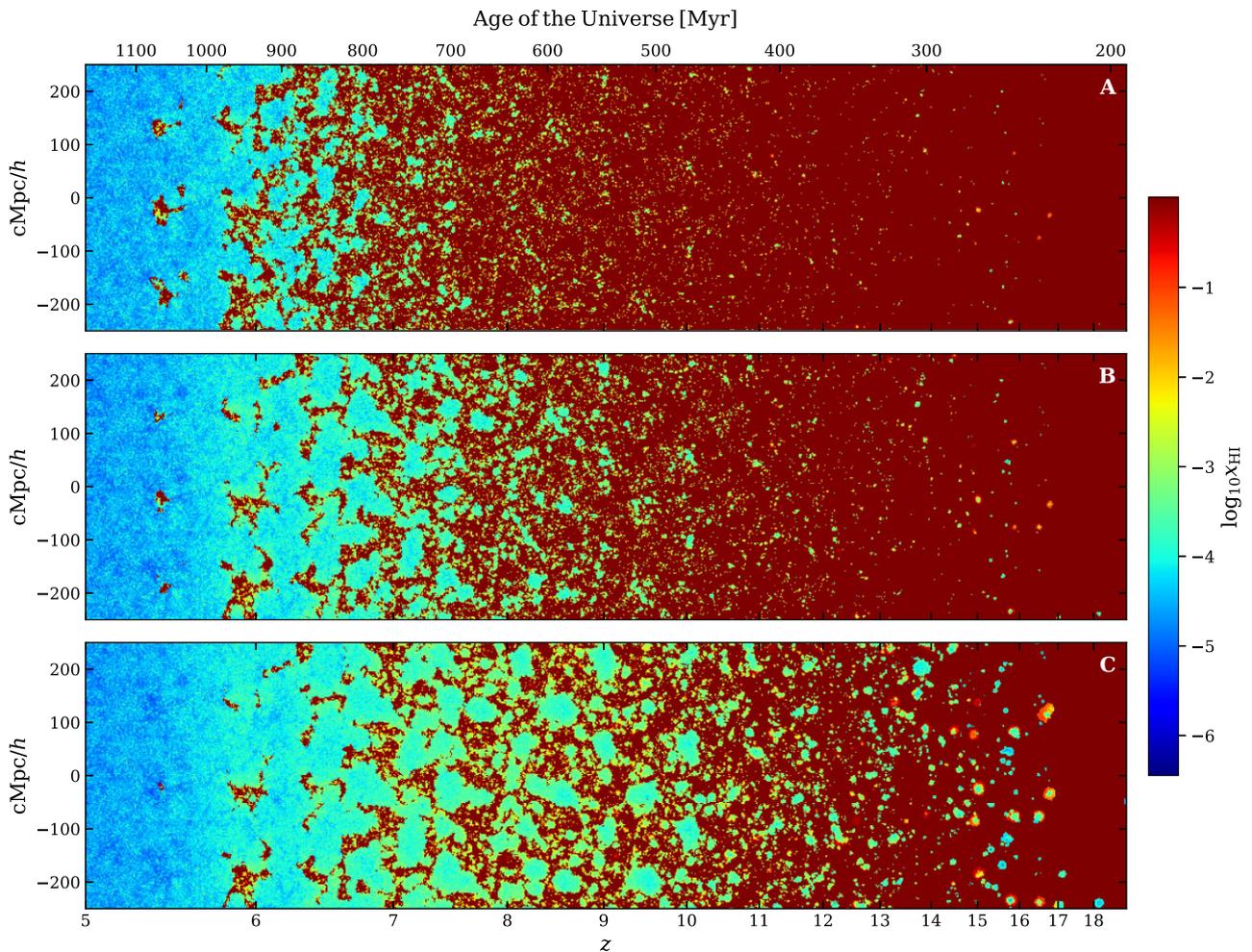

**Figure 7.** Lightcones of the neutral hydrogen fraction for our fiducial (panel a), the 'Early' (panel b), and 'Extremely Early' (panel c) reionization histories.

The observed distribution is shown by the black curve, while the fiducial, 'Early', and 'Extremely Early' models are shown by the dark blue, red, and yellow curves, respectively. Starting with the highest redshift bin, the models agree reasonably well with the observed distribution. The effective optical depth in ionized regions (low $\tau$) is in slightly better agreement with the data in the fiducial model compared to the other two. The effective optical depth in the neutral regions (high $\tau$), on the other hand, is fit better by the 'Early' and 'Extremely Early' models. The fiducial model, with its higher volume-averaged neutral fraction, possesses regions that are more neutral compared to the other models. Progressing to the redshift bin of $5.83 < z < 6$, the agreement at the tail end of the effective optical depth distribution in the fiducial model is also not perfect, suggesting slightly more neutral gas than the observed distribution. The 'Early', and 'Extremely Early models, in contrast, agree here better with the data except at the largest optical depth, where they fall short in predicting the full range of the high optical depth tail. Note, however, that the simulations snapshots contributing to the highest redshift bins are at $z = 6.14$ and $z = 5.94$, towards the upper end of redshifts probed by the observations in these two redshift bins. In the subsequent redshift bins, where outputs from our models are more central in the redshift bin, the fiducial model reproduces the observed distribution reasonably well. On the other hand in the 'Extremely Early' model the distribution falls somewhat short at high effective optical depth, compared to the data. In the simulations large effective optical depths in this redshift range correspond to the last remaining neutral islands and the model appears to produce too few of these in the redshift range $5.5 < z < 5.83$ to be consistent with the data. The effective optical depth distribution in the 'Early' model lies in between the other two cases. Below $z \sim 5.4$, the volume-averaged neutral fraction in the 'Extremely Early' model is significantly lower compared to the fiducial and the 'Early' model despite the effective optical depth CDFs in all three models being similar. This is due to a small difference in the redshift when the last neutral island disappear and reionization fully completes. Note also, that the neutral hydrogen fraction of the 'Extremely Early' model in this redshift range is also significantly lower than the possible range of values found by Gaikwad et al. (2023). As the 'Extremely Early' model fits the effective optical depth distribution, this would suggest that the error bars in Gaikwad et al. (2023) are somewhat underestimated in the redshift range where the last neutral regions are ionized and the volume-averaged neutral fraction changes rapidly.

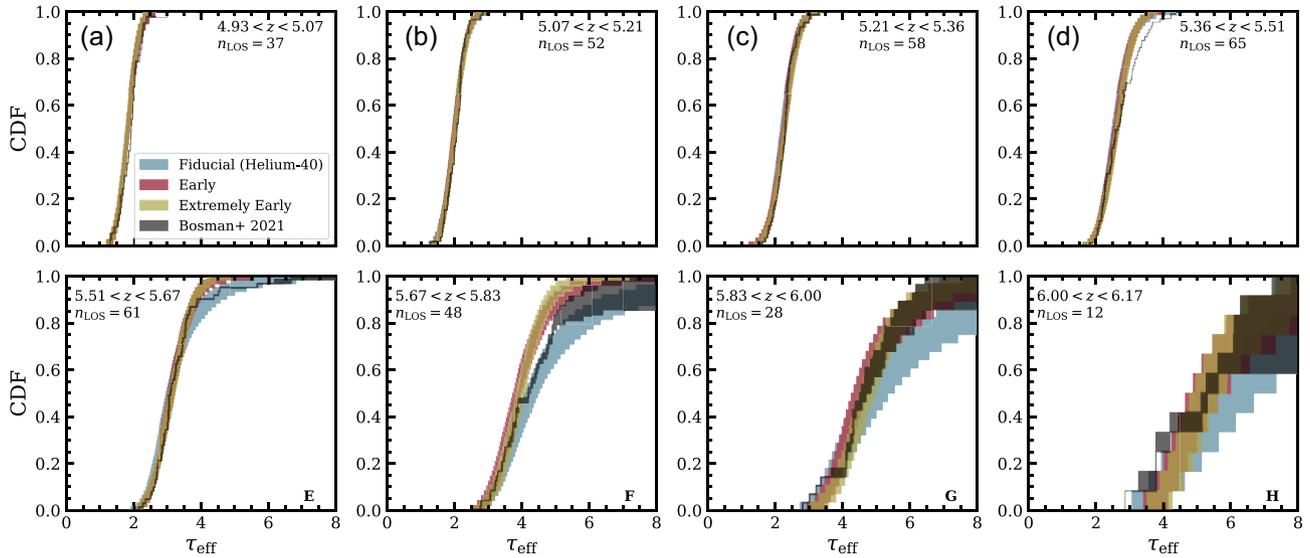

**Figure 8.** Evolution of the CDFs of the effective Ly α opacity in the fiducial (blue), 'Early' reionization (red), and 'Extremely Early' reionization (yellow) models in narrow redshift bins from $z \sim 6$ to 5. Also shown are the CDFs measured by Bosman et al. (2022). Each panel shows the number of lines of sight $n_{\rm LOS}$ observed by Bosman et al. (2022) in that redshift bin. The simulated samples are chosen have the same size as the observed sample in the respective redshift bin. Shaded regions show the range between the 15.87 and 84.13 percentiles.

Scrutinizing the 'Extremely Early' and 'Multi-H-70' models more closely reveals another subtle aspect of reionization history impact on effective optical depth distributions. Despite these models exhibiting almost identical volume-averaged neutral fractions in the redshift range $5.6 \lesssim z \lesssim 6$, as shown in Fig. 5(b), their CDF of effective optical depth differ within the range $5.67 < z < 5.83$, as shown in Fig. 4.

A comparison of our results with the data by Bosman et al. (2022) reveals the quality of these newer data compared to previous measurements. The narrower distribution, particularly at higher optical depths, makes it more difficult for the simulations to align with the full CDF. Past work presented by Keating et al. (2020a) and Kulkarni et al. (2019a), managed to easily achieve a fit to the observed CDF within the errors of previous measurements by calibrating to the mean flux. We found this approach more challenging when using the significantly tighter constraints of the new measurements by Bosman et al. (2022).

The disparities in the distribution of neutral and ionized regions across the models naturally lead us to look into the spatial configuration and characteristics of the ionized bubbles within these simulations. Given the recent data from the *JWST* that sheds new light on the relation of ionized bubbles and galaxies, it becomes pertinent to examine the spatial distribution of bubbles and ionizing sources in our simulations with different reionization histories. We will do this in the following section.

## 6 MORPHOLOGY AND SIZES OF IONIZED BUBBLES

In the previous section, we saw that a wide range of reionization histories are consistent with the observed Ly α opacity distribution at $5 \lesssim z \lesssim 6.2$. As expected and as is visually apparent the different reionization histories differ strongly when the first ionized regions appear and grow into large ionized bubbles. As already discussed above, the transmission of Ly α emission from reionization-epoch galaxies depends strongly on whether they are affected by the damping wings of intervening neutral gas along the line of sight to the observer (Umeda et al. 2024). Modelling this is complex as the Ly α transmission depends on the intrinsic systemic redshift and width of the Ly α emission line. It also depends on the peculiar velocity between galaxy and surrounding neutral gas as well as the distance of the galaxy to the bubble 'wall'. Furthermore, the neutral hydrogen column density of the bubble wall and the absorption by residual neutral hydrogen within the ionized bubble (Weinberger et al. 2018; Keating et al. 2024) also plays a role. To bypass the many uncertainties in this modelling, the size of the ionized bubble in which the reionization-epoch galaxy is located in is often taken as a proxy for high Ly $\alpha$ transmission. Canonically a radius of one proper Mpc is assumed to ensure high Ly $\alpha$ transmission (Weinberger et al. 2018). In this section, we have therefore quantified the bubble growth in our simulation and calculated the evolution of the fraction $N_{h,frac}$ of our ionizing sources as characterized by their total host halo mass that sit in ionized bubbles of a given size. We will see that $N_{h,frac}$ rises to unity significantly earlier than the volume averaged neutral hydrogen fraction $\langle x_{HI} \rangle_v$ due to the strongly biased clustering of haloes sufficiently massive to host ionizing sources at $z > 6$.

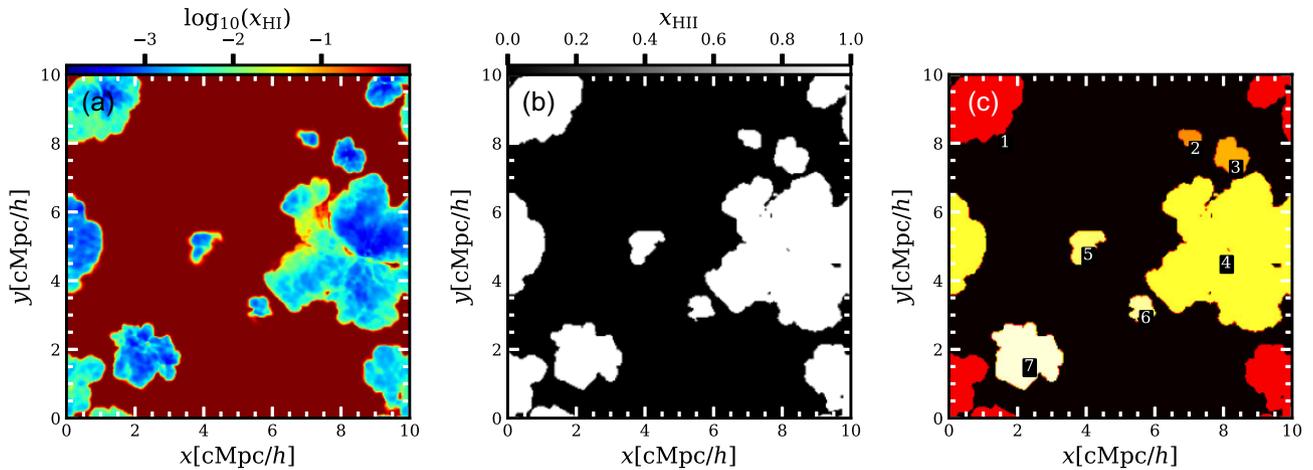

**Figure 9.** An illustration of how ionized hydrogen bubbles are identified in our simulations. The left panel shows a small, 10 Mpc $h^{-1}$ wide two-dimensional slice from the neutral hydrogen distribution in the simulation box. This information is converted into a binary, black-and-white map in the central panel, with ionized regions ($x_{HI} < 0.5$) shown in white and neutral regions ($x_{HI} > 0.5$) shown in black. In the right panel, a friends-of-friends algorithm now identifies distinct bubbles. Each bubble is numbered. In this example, we see that there are seven bubbles, taking into account the periodic boundary conditions.

### 6.1 Calculating bubble sizes

In Fig. 9, we demonstrate how we measure bubble sizes within our simulation, leveraging image processing techniques. The left-hand plot of the figure provides a visual representation of the neutral hydrogen distribution within a two-dimensional cross-section of our 10 cMpc $h^{-1}$ simulation box. A smaller simulation box was selected to allow visual verification of the bubble count. Here, ionized regions are shown in blue, while more neutral areas are presented in red. For the bubble identification in our simulations, we first convert the colour image depicting neutral hydrogen distribution into a binary black and white version, as illustrated in the middle panel of the figure. This binary representation requires defining a threshold to distinguish between ionized and neutral regions. We set a threshold value of 0.5 for the neutral hydrogen fraction (Chardin, Aubert & Ocvirk 2012a, 2014), above which a cell is categorized as neutral, while those below are treated as ionized. During this conversion ionized regions become white, while neutral regions are transformed into black. The next step involves applying a connected component labelling algorithm (CCL) to count and identify each ionized (white) region as a distinct bubble. This process identifies contiguous regions within a binary image where pixel values match. We then use the python routine `scipy.measure.label`, with a setting `connectivity = 3` that ensures that a voxel connects to its neighbours if they share any common points, to implement the CCL algorithm. However, given the periodic nature of our simulation volume, the standard algorithm overlooks bubbles that extend across the boundaries of the simulation volume. To rectify this we exploit the periodicity of our simulations and link such regions across the simulation boundaries. In the right-hand panel, we show the implementation of the modified CCL algorithm where each bubble has a unique colour, and number to indicate its count. The methodology applied in our simulation successfully identifies seven distinct bubbles. A particularly noteworthy example is bubble 1, highlighted in bright red, located at the left edge of the plot. This bubble shows the effect of the periodic boundary conditions, extending across both the left-to-right and top-to-bottom edges of the simulation box. This periodicity is accurately captured by our technique, underscoring its effectiveness in identifying and quantifying the dimensions of ionized bubbles within the simulation. This approach can be straightforwardly extended to a three-dimensional analysis. Once the bubbles are identified, we can quantify the number of cells within each bubble, which corresponds to its volume $V$. Assuming a spherical shape for the bubbles, we can then infer their 'radius' $R$ using the relation $V = 4\pi R^3/3$. For a simulation cube measuring 160 cMpc $h^{-1}$, this method yields a maximum radius of 99.2 cMpc $h^{-1}$. Although the assumption of a spherical bubble is not entirely accurate, it does provide a reasonable estimate of the radius of the bubble radius for our purposes. Other bubble measuring techniques are the 'Distance transform' (Zahn et al. 2007, 2011) and 'mean free path' method (Mesinger & Furlanetto 2007), which assign a 'radius' to each cell in the simulation. Other methods to identify ionized bubbles employ the friends-of-friends algorithm (Iliev et al. 2006; Chardin, Aubert & Ocvirk 2012b, 2014), while others employ the watershed algorithm which also uses the CCL algorithm (Soille 2003; Lin et al. 2016).

### 6.2 Bubble sizes and the timing of reionization

The lightcones depicted in Fig. 7 offer a visual illustration of the growth of ionizing bubbles for the three different reionization histories. A quantitative assessment of these structures involves calculating bubble sizes and examining their distribution across the simulation volume. Employing the methodology outlined above, we apply this analysis to the entire three-dimensional simulation box. The outcomes are presented in Fig. 10.

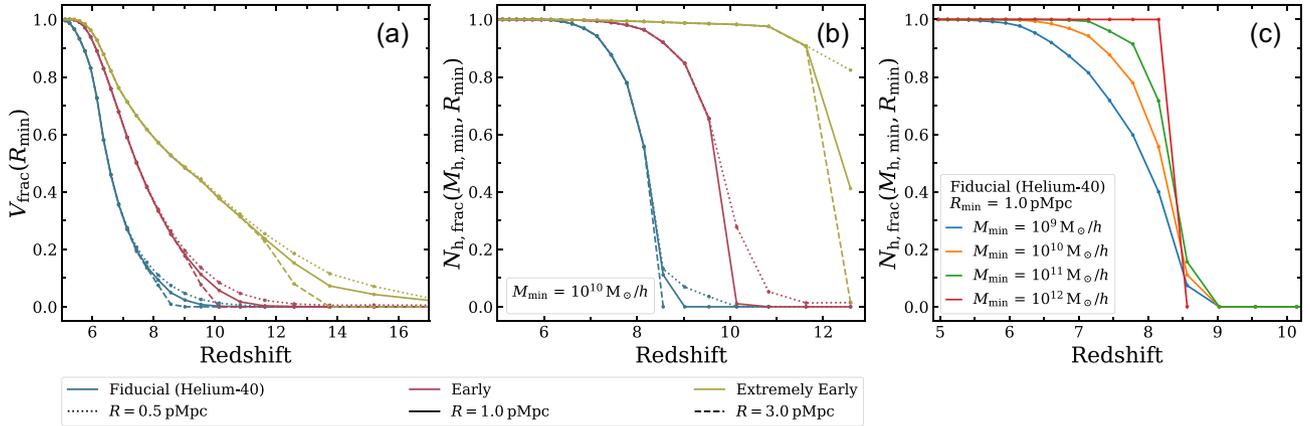

**Figure 10.** Panel (a) shows the fraction of volume in ionized bubbles ($x_{HI} < 0.5$) with radius greater than 0.5 pMpc (dotted curves), 1 pMpc (solid curves), and 3 pMpc (dashed curves) in our fiducial, 'Early', and 'Extremely Early' reionization models. For these three bubble sizes, Panel (b) shows the fraction of haloes with mass greater than $10^{10}$ $M_\odot\,h^{-1}$ that are located in the ionized regions in these three models. Panel (c) focuses on our fiducial model, and a minimum bubble radius $R_{min}$ of 1 pMpc, for which this panel shows the number of haloes with mass greater than $10^9$, $10^{10}$, $10^{11}$, and $10^{12}$ $M_\odot\,h^{-1}$ that are located in the ionized regions.

The left plot in Fig. 10 presents a comparative analysis of the evolution of the fraction of the volume ($V_{frac}$) occupied by ionized bubbles exceeding a certain minimum radius with redshift for different simulation models. The plot categorizes bubbles based on their size: dotted curves represent bubbles with a minimum radius of 0.5 pMpc, solid curves are for 1 pMpc, and dashed curves are for 3 pMpc. Each line colour corresponds to a different simulation model. At higher redshifts, a larger proportion of the simulation volume is occupied by smaller bubbles with radius 0.5 pMpc. However, as the universe evolves, these smaller bubbles quickly merge or expand, leading to a convergence in the volume occupied by bubbles of different sizes. The timing of reionization onset in each model plays a crucial role in determining the bubble size evolution. For instance, the 'Extremely Early' model, in which reionization starts earliest, exhibits a remarkable 30 per cent of its total volume occupied by ionized bubbles with size $\gtrsim$ 1 pMpc as early as $z = 11.5$. In contrast, the 'Early' and other models show negligible volume fractions occupied by ionized regions with size $\gtrsim$ 1 pMpc at this redshift. By $z = 10$, the 'Early' model shows a substantial increase, with around 10 per cent of its volume containing ionized bubbles $\gtrsim$ 1 pMpc. Overall, these trends closely mirror the evolution of the volume-averaged ionization fraction for each simulation.

In the middle plot of Fig. 10, we examine the fraction $N_{h,frac}$ of haloes with a mass exceeding $10^{10}$ $M_\odot\,h^{-1}$ that are located within ionized regions of a radius greater than a specified minimum, $R_{min}$. The colour coding and line styles in this plot are consistent with those in the left plot. When considering the redshift evolution of sources and bubbles we should mention that we post-process simulation snapshots taken at 40 Myr intervals which requires some attention. Consider at redshift $z_1$ sources emit photons, leading to the formation of ionized bubbles. By redshift $z_2$, the source information is updated based on data from the hydrodynamical simulations, changing the source locations. When identifying sources within bubbles we take the bubble sizes at redshift $z_2$ and the source position at redshift $z_1$, i.e one snapshot prior to the redshift at which the bubble sizes are calculated. Looking at the plots we can see a rapid convergence of the curves for all bubble sizes considered, mirroring the trends observed in the volume fraction analysis. Furthermore, the fraction of haloes found within ionized bubbles of size $\gtrsim$ 1 pMpc escalates swiftly towards unity, indicating that shortly after the onset of reionization, the majority of these haloes are located within ionized regions. Focusing on the fiducial 'Helium-40' model as an example, we observe that by around $z \sim 7$, all the haloes in this category are situated within ionized bubbles greater than 1 pMpc in radius. Yet, at this same redshift, these large bubbles constitute only about 40 per cent of the total volume of the simulation. This pattern is consistent across all models, underscoring a strong correlation between the locations of the haloes and the ionized regions. This correlation suggests that the formation and expansion of ionized bubbles during the reionization process are closely linked to the presence of these massive haloes. As these haloes are likely to be significant sources of ionizing radiation, their spatial distribution plays a crucial role in shaping the reionization morphology, leading to a scenario where these haloes are often found within at the centre of the ionized regions.

Fig. 10(c) shows the distribution of haloes of varying masses within ionized regions of radius > 1 pMpc during the reionization process for the fiducial model. This plot is particularly insightful for understanding how halo mass correlates with their likelihood of being within large ionized bubbles. There is a clear trend indicating that haloes with greater mass are more rapidly encompassed by ionized regions. This is exemplified by the curves representing different mass cut-offs, where the heavier the halo mass (ranging from $10^9$ to $10^{12}$ $M_\odot\,h^{-1}$), the sooner it is found within an ionized bubble. Note that for the halo mass of $10^{12}$ $M_\odot\,h^{-1}$ the curve starts at $z = 8.56$ as haloes of this mass are not present in the simulation above this redshift. Furthermore, we see that as reionization nears its end, even lower mass haloes (as indicated by the blue curve for haloes with mass around $10^9$ $M_\odot\,h^{-1}$) are eventually found within large ionized regions. This inclusion reaches completion around $z \sim 5.5$, at which point the ionized bubbles occupy the entirety of the simulation volume, as depicted in the volume fraction shown in the left plot.

Our analysis indicates that while ionized bubbles of various sizes are present at the onset of reionization, they quickly merge with one another. The timing of the reionization onset plays a crucial role in determining when these bubbles will encompass the entire simulation volume. We also find that more massive haloes have a high probability to be found within large ionized regions shortly after the start of reionization. This suggests that these haloes, due to their higher ionizing output, are mainly responsible for shaping the early structure of ionized regions. By the end of reionization, haloes across the mass spectrum, including the low-mass ones, are integrated into the ionized volume. Our fiducial model seems to not have any number of high mass haloes in ionized bubbles with radius $> 0.5$ Mpc at $z = 10$, which may be in contrast with recent *JWST* results claiming detections of bright Ly $\alpha$ emitters already at $z \sim 10$ and beyond. We will discuss the implications for the optical depth for Thomson scattering of the CMB in the next section.

# 7 DISCUSSION

In our simulations, we explored the effect of a range of parameters including the number of frequency bins in the radiative transfer computation, the source SED, helium chemistry, early stages of reionization, and the source models on several observables related to the EoR. Some of these parameters, in varying combinations, have also been subjects of investigation in previous studies. It is thus instructive to compare our findings with the literature.

Our work builds upon the foundations laid by Kulkarni et al. (2019a) and Keating et al. (2020b), who utilized the ATON code for monofrequency hydrogen simulations, calibrating them to the Ly $\alpha$ mean flux data from Bosman et al. (2018). Notably, Keating et al. (2020b) explored scenarios with elevated Thomson optical depth $\tau_{CMB}$ values. We improve on these simulations by calibrating them to the more recent and stringent Ly $\alpha$ flux constraints provided by Bosman et al. (2022), which place tighter bounds on the distribution of regions with large Ly $\alpha$ optical depth. Our models with higher $\tau_{CMB}$ values, labelled 'Early' and 'Extremely Early', probe the morphological characteristics of H II regions at higher redshifts than the previous studies based on ATON simulations. We have also explored different source populations namely the 'Democratic' and 'Oligarchic' models.

Our findings with regard to the effect of including helium echo those of Ciardi et al. (2012), especially regarding the observation that the inclusion of helium can postpone the reionization process and lead to diminished temperatures in simulations including helium relative to those considering only hydrogen, particularly at redshifts $z \gtrsim 10$. A significant advantage of our approach is the considerable scale of our simulations, which utilize a box size nearly fivefold larger than that employed by Ciardi et al. (2012) and our simulations are calibrated to observed Ly $\alpha$ forest data. The THESAN simulation suite, presented by Kannan et al. (2021), adopts a coupled radiation-hydrodynamics approach including helium with multiple frequency bins. However, their fiducial run, as analysed by Garaldi et al. (2022), does not agree as closely with the E-XQR-30 data from Bosman et al. (2022) as our simulations do. Additionally, the THESAN suite exhibits a monotonically increasing emissivity, compared to a dip required by our models, a characteristic that Cain et al. (2024) attributes to the implementation of a reduced speed of light (see also Qin et al. 2021). Note here that in our models the dip is less pronounced in the Oligarchic, 'Early' and 'Extremely Early' models than in our fiducial model.

Cain et al. (2024) scrutinized the influence of 'Oligarchic' versus 'Democratic' source models with ray-tracing simulations, finding the 'Oligarchic' model less favoured due to its higher neutral fraction. Conversely, Garaldi et al. (2022) introduced the 'THESAN-LOW-2' model, where sources above $10^{10}$ M$_\odot$ contribute to ionizing radiation, but noted that such a model has a consistently lower neutral fraction than their fiducial models. Our analysis, however, reveals that adjusting the ionizing volume emissivity permits the calibration of an 'Oligarchic' model that closely tracks the 'Democratic' model in terms of its ionization history but also fits the Ly $\alpha$ effective optical depth CDF.

Neyer et al. (2024) and Lu et al. (2024) recently used the THESAN simulations and the 21CMFAST framework (Mesinger, Furlanetto & Cen 2011), respectively, to investigate the relation of ionized regions and ionizing sources. Similar to previous studies they found that galaxy over-densities are more likely to be found in ionized regions, and that brighter galaxies are surrounded by larger ionized bubbles. We have followed on from this here and have investigated the spatial distribution of ionizing sources within the ionized regions in our models consistent with observations of the Ly $\alpha$ forest.

Park et al. (2021) looked at the Ly $\alpha$ transmission from galaxies in the coupled radiative hydrodynamical simulation CoDa II (Ocvirk et al. 2020) and found that bright galaxies are less affected by reionization. This has also been found in several observational studies. Endsley & Stark (2022) found that nine out of ten galaxies show Ly $\alpha$ emission at $z = 6.8$. Harish et al. (2022) found 15 Ly $\alpha$ emitters (LAEs) out of a sample of 21. These observation agree with predictions from our reionization models, as illustrated in Figs 10(b) and (c), showing that, across various timing models and halo masses, more than 85 per cent of sources are situated within ionized bubbles larger than 1 pMpc at $z = 7$. This bubble size is large enough for the Ly $\alpha$ emission to sufficiently redshift before it encounters the bubble wall, allowing into to escape without resonant scattering (Weinberger et al. 2018). Moreover, the study by Endsley et al. (2021) indicated a relatively constant transmission rate of Ly $\alpha$ emission from massive galaxies between $z \sim 6$ and 7, contrasting with the significant decline observed in less massive galaxies, as reported by Fuller et al. (2020). Our models mirror this behaviour, as evidenced in Fig. 10(c), where we see that larger haloes are incorporated into ionized zones much earlier than their smaller counterparts. Consequently, between $z \sim 6$ and 7, there is an observed increase of about 20 per cent in the proportion of low-mass ($10^9$ M$_\odot$ $h^{-1}$) sources exhibiting Ly $\alpha$ emission, in stark contrast to a marginal change for haloes with masses exceeding $10^{10}$ M$_\odot$ $h^{-1}$.

The detection of Ly $\alpha$ emission at high redshifts, where the IGM becomes increasingly neutral, provides important insights into the early stages of reionization. Pentericci et al. (2018), and Harish et al. (2022) have identified galaxies within the redshift range of approximately 6–7, while studies by Oesch et al. (2015), Stark et al. (2016), Jung et al. (2019), Tilvi et al. (2020) and Jung et al. (2020) have extended these findings to galaxies around $z \sim 7.5$. Furthermore, Zitrin et al. (2015) reported a Ly $\alpha$ emitter at $z = 8.68$, and Bunker et al. (2023) discovered an LAE at an even more distant redshift of $z = 10.60$. In the context of our simulations, particularly the fiducial model, we observe that up to $z \sim 8$, haloes with masses exceeding $10^9$ M$_\odot$ $h^{-1}$ have a 50 per cent chance of residing within ionized bubbles large enough to facilitate the escape of Ly$\alpha$ emissions without encountering a large optical depth to resonant scattering. This probability diminishes sharply past $z \sim 9$, challenging the viability of our fiducial reionization history in light of the recent *JWST* observations of LAEs at $z \gtrsim 9$. To address this, our 'Early' and 'Extremely Early' have reionization histories where ionized bubbles of sufficient size exist at $z \sim 10.50$, to allow for a high transmissivity of Ly $\alpha$ emission. Assessing this more quantitatively will need larger observed samples and more detailed modelling of the Ly $\alpha$ emission and will depend in particular on the redshift of the Ly $\alpha$ emission before it is scattered by the circumgalactic medium (CGM) and the IGM (see Sadoun, Zheng & Miralda-Escudé 2017 and Weinberger et al. 2018 for a discussion of the combined effect of CGM and IGM on Ly $\alpha$ transmissivity and Tang et al. 2024 for a recent summary of the observed properties of Ly$\alpha$ emitters at $z \sim 5$–6 and a discussion of implications for Ly$\alpha$ transmissivity at earlier redshifts). The modelling will be further complicated by an possible increased contribution from AGN to the Ly$\alpha$ emission (Maiolino et al. 2023; Scholtz et al. 2023).

In summary, the expected strong biasing of ionizing sources and the implied tendency to reside early on in rather large ionized bubbles helps to explain the presence of bright LAEs at redshifts where the Universe is still predominantly neutral. However, we also note again that the 'Extremely Early' model exhibits a $\tau_{\rm CMB}$ which is in $> 3\sigma$ in tension with measurements reported by Planck Collaboration VI (2020).

## 8 CONCLUSIONS

In this paper, we have studied a range of reionization histories consistent with the observed mean Ly$\alpha$ transmission and the Ly$\alpha$ opacity distributions in the E-XQR-30 sample at $z > 5$. We perform cosmological simulations post-processed for radiative transfer using ATON-HE, an updated version of the ATON code that includes helium.

We calibrate multiple radiative transfer simulations of reionization to the mean Ly$\alpha$ transmission $\langle F \rangle$ reported in the new measurements of the Ly$\alpha$ opacity by Bosman et al. (2022). The new Ly$\alpha$ opacity measurements are consistent with previous studies, but have significant lower uncertainties with CDFs that show a somewhat less pronounced tail at high optical depth. We notice that the values of the neutral hydrogen fraction that set the mean flux $\langle F \rangle$ are different from the values that dominate the volume-averaged neutral hydrogen fraction, $\langle x_{\rm HI} \rangle_v$. Consequently, models calibrated to the $\langle F \rangle$ measurements reported by Bosman et al. (2022) can still have some difference in $\langle x_{\rm HI} \rangle_v$ at these redshifts. The less pronounced tail at high optical depth thereby appears to require a somewhat smaller volume filling factor of remaining neutral islands than was suggested by previous measurements of the Ly$\alpha$ opacity.

Relative to our previous monofrequency simulations, multifrequency simulations calibrated to the mean Ly$\alpha$ transmission $\langle F \rangle$ show expected changes. The gas temperature at mean density in these simulations is higher than in their monofrequency counterparts. The required photon emissivity is lower. There is no significant change in the evolution of the volume-averaged neutral hydrogen fraction, $\langle x_{\rm HI} \rangle_v$, or the mean free path. Next, adding helium to our simulations and continuing to enforce calibration to the mean Ly$\alpha$ transmission by adjusting the ionizing volume emissivity somewhat reduces the gas temperature for a given source spectrum. At the same time the required ionizing volume emissivity increases to provide the necessary photons to ionize He I to He II. The mean free path also decreases by a small amount. For accurate quantitative inference on the temperature evolution of the IGM and how it relates to source SEDs especially if harder sources like QSOs are involved and/or if one is explicitly interested in the ionization state of helium, our modifications of ATON are obviously needed. If one is mainly interested in how hydrogen reionization proceeds, hydrogen-only, mono-frequency simulations require significantly fewer resources but as discussed have obvious shortcomings.

We also study the effect of the ionizing source model on reionization. In our fiducial model the ionizing photon emissivity $\dot{N}_{\rm ion}$ of a source is proportional to its total halo mass $M$. We consider a Democratic model in which $\dot{N}_{\rm ion}$ is independent of $M$, and an Oligarchic model in which $\dot{N}_{\rm ion}$ is also proportional to $M$ but is non-zero only in haloes with the highest masses. These three models display interesting differences that can be interpreted as resulting from the correlation between the IGM density and halo locations. In order to reproduce the observed mean Ly$\alpha$ flux $\langle F \rangle$, the Oligarchic model requires a somewhat smaller emissivity while the Democratic model requires a greater emissivity relative to the fiducial model. In the Oligarchic model, reionization starts slightly earlier than in the fiducial model and ends slightly later. In the Democratic model, the early stages of reionization are very similar to those in the fiducial model, but the end of reionization occurs also slightly later than in the fiducial model. All three models are broadly consistent with the distribution of the Ly$\alpha$ opacity and overall the differences in this regard are surprisingly small.

The Ly$\alpha$ opacity measurements from the Ly$\alpha$ forest data still leave substantial freedom for the evolution of the reionization in its early stages. We have considered here two model variations to investigate this. While our fiducial simulation has its midpoint of reionization at $z = 6.5$, our 'Early' reionization model has its midpoint at $z = 7.5$ and our 'Extremely Early' reionization model has its midpoint at $z = 9.5$. All three of these models are calibrated to the observed mean Ly$\alpha$ transmission. They are also broadly in agreement with the measured Ly$\alpha$ opacity distributions. The 'Early' and 'Extremely Early' models thereby complete reionization slightly earlier than the fiducial model. Note, however, that these differences are again small. The redshift at which $\langle x_{\rm HI} \rangle_v = 0.1$ per cent varies between $z = 5.2$ and 5.4.

We have quantified the size of ionized 'bubbles' in our simulations using an extension to the connected-CLA. As expected the fraction of the simulation volume occupied by bubbles of a minimum size tracks the progress of reionization. The spatial distribution of haloes with masses able to host bright galaxies at high redshift ($M > 10^{10}$ M$_\odot$ $h^{-1}$) is highly biased. In all our models, the fraction of such haloes located in bubbles with sizes greater than 0.5 pMpc rapidly approaches unity in the early stages of reionization. The fraction of such haloes located in bubbles with sizes greater than 1 and 3 pMpc also rises significantly earlier than the volume-averaged neutral fraction. However, only in our 'Extremely Early' model this happens already at $z \gtrsim 10$. In particular our fiducial model and perhaps also our 'Early' model might thus be difficult to reconcile with the recent *JWST* detections of abundant Ly$\alpha$ emission from galaxies up to $z \sim 10$ and beyond. As sample sizes increase, the occurrence of Ly$\alpha$ emission at the highest redshifts should become an increasingly important constraint for inferring the early stages of reionization from *JWST* data and may already now be in (mild) tension with the lower end of values for the Thomson optical depth suggested by the Planck data.

## ACKNOWLEDGEMENTS

We thank Prakash Gaikwad, Vid Iršič, Margherita Molaro, Ewald Puchwein, and Sindhu Satyavolu for helpful discussions. We also thank the referee, Nick Y. Gnedin, for their very helpful comments. This work was performed using resources provided by the Cambridge Service for Data Driven Discovery (CSD3) operated by the University of Cambridge Research Computing Service (www.csd3.cam.ac.uk), provided by Dell EMC and Intel using Tier-2 funding from the Engineering and Physical Sciences Research Council (capital grant EP/T022159/1), and DiRAC funding from the Science and Technology Facilities Council (www.dirac.ac.uk). The project was also supported by a grant from the Swiss National Supercomputing Centre (CSCS) under project ID s1114. Support by ERC Advanced Grant 320596 'The Emergence of Structure During the Epoch of Reionization' is gratefully acknowledged. MGH has been supported by STFC consolidated grants ST/N000927/1 and ST/S000623/1. GK gratefully acknowledges support by the Max

Planck Society via a partner group grant. GK is also partly supported by the Department of Atomic Energy (Government of India) research project with Project Identification Number RTI 4002. Part of the work has been performed as part of the DAE-STFC collaboration Building Indo-UK collaborations towards the Square Kilometre Array' (STFC grant reference ST/Y004191/1). SA also thanks the Science and Technology Facilities Council for a PhD studentship (STFC grant reference ST/W507362/1), and the University of Cambridge for providing a UKRI International Fees Bursary.

## DATA AVAILABILITY

The simulation data used in this work is available on reasonable request.

## APPENDIX A: RADIATIVE TRANSFER AND THERMOCHEMISTRY IN ATON-HE

This section covers the key features of the original code, as well as the modifications we implemented. We followed the description in Aubert & Teyssier (2008) and Rosdahl et al. (2013).

### A1 Moments of the radiative transfer equation

The radiative transfer equation in comoving coordinates is given by

$$-\kappa_\nu I_\nu + \eta_\nu = \frac{1}{c}\frac{\partial I_\nu}{\partial t} + \frac{\hat{n}.\nabla I_\nu}{a_{em}} + \frac{3H(t)I_\nu}{c} - \frac{H(t)}{c}\nu\frac{\partial I_\nu}{\partial \nu}, \quad (A1)$$

where the second term describes the propagation of radiation while the $1/a$ factor takes into account the cosmic expansion. The third term is the effect due to cosmological redshifting, and the last term describes the dilution of radiation (Gnedin & Ostriker 1997). The quantity $I_\nu(\boldsymbol{x}, \boldsymbol{n}, t)$ is specific intensity of the radiation, $\kappa_\nu(\boldsymbol{x}, \boldsymbol{n}, t)$ is the absorption coefficient, and $\eta_\nu(\boldsymbol{x}, \boldsymbol{n}, t)$ is the source function. Note that $a_{em}$ is approximately 1 for the short photon crossing time and therefore we can write,

$$-\kappa_\nu I_\nu + \eta_\nu = \frac{1}{c}\frac{\partial I_\nu}{\partial t} + \hat{n}.\nabla I_\nu + \frac{3H(t)I_\nu}{c} - \frac{H(t)}{c}\nu\frac{\partial I_\nu}{\partial \nu}. \quad (A2)$$

Before moving on to the moment equations, we define the radiation energy density (Rosdahl & Teyssier 2015) as

$$E_\nu = \frac{1}{c}\int I_\nu(\boldsymbol{x}, \boldsymbol{n}, t)\mathrm{d}\Omega. \quad (A3)$$

We define the radiation flux as

$$F_\nu^i = \int n^i \frac{I_\nu(\boldsymbol{x}, \boldsymbol{n}, t)}{h\nu}\mathrm{d}\Omega, \quad (A4)$$

and the radiation pressure as

$$P_\nu^{ij} = \frac{1}{c}\frac{1}{h\nu}\int n^i n^j I_\nu(\boldsymbol{x}, \boldsymbol{n}, t)\mathrm{d}\Omega. \quad (A5)$$

For the zeroth-order moment equation we integrate equation (A2) over the solid angle. All angular integrations are over the whole sphere. This leads to

$$-\int \kappa_\nu I_\nu d\Omega + \int \eta_\nu d\Omega = \int \frac{1}{c}\frac{\partial I_\nu}{\partial t}d\Omega + \int \hat{n}.\nabla I_\nu d\Omega$$
$$+ \int \frac{3H(t)I_\nu}{c}d\Omega - \int \frac{H(t)}{c}\nu\frac{\partial I_\nu}{\partial \nu}d\Omega. \quad (A6)$$

Using the definition of the radiation energy density (equation A3), we get

$$-c\kappa_\nu E_\nu + \int \eta_\nu d\Omega = \frac{\partial E_\nu}{\partial t} + \int n^i.\frac{\partial I_\nu}{\partial x^j}d\Omega$$
$$+ 3H(t)E_\nu - H(t)\nu\frac{\partial E_\nu}{\partial \nu}. \quad (A7)$$

Substituting equation (A4),

$$-c\kappa_\nu E_\nu + \int \eta_\nu d\Omega = \frac{\partial E_\nu}{\partial t} + h\nu\nabla\vec{F}_\nu + 3H(t)E_\nu - H(t)\nu\frac{\partial E_\nu}{\partial \nu}, \quad (A8)$$

and dividing by the photon energy we get,

$$-c\kappa_\nu N_\nu + S_\nu = \frac{\partial N_\nu}{\partial t} + \nabla\vec{F}_\nu + 3H(t)N_\nu - \frac{1}{h\nu}H(t)\nu\frac{\partial E_\nu}{\partial \nu}. \quad (A9)$$

The variation of energy with the frequency is ignored and we are left with

$$-c\kappa_\nu N_\nu + S_\nu = \frac{\partial N_\nu}{\partial t} + \nabla \vec{F}_\nu + 3H(t)N_\nu - \frac{1}{h\nu}H(t). \quad (A10)$$

To get the first-order moment equation, we multiply equation (A2) with $\vec{n}$, divide by $h\nu$, and integrate to get

$$\int (-\kappa_\nu I_\nu + \eta_\nu)\frac{n^i}{h\nu}d\Omega = \int \left[\frac{n^i}{c}\frac{1}{h\nu}\frac{\partial I_\nu}{\partial t} + n^k n^i \frac{1}{h\nu}\frac{\partial I_\nu}{\partial x^j}\right]d\Omega$$
$$+ \int \frac{3n^i H(t)I_\nu}{c}\frac{1}{h\nu}d\Omega - \int \frac{H(t)}{c}\frac{1}{h\nu}\nu\frac{\partial I_\nu}{\partial \nu}. \quad (A11)$$

Substituting the definition for radiation flux (equation A4) and radiation pressure (equation A5),

$$-c\kappa_\nu \vec{F}_\nu + \int c\eta_\nu n^i d\Omega = \frac{\partial \vec{F}_\nu}{\partial t} + c^2\frac{\partial P_\nu}{\partial x^j} + 3H(t)\vec{F}_\nu$$
$$- \frac{H(t)}{h}\int \left(\frac{I_\nu}{\nu} + \frac{\partial (I_\nu/\nu)}{\partial \nu}\nu\right). \quad (A12)$$

Simplifying the integral,

$$-c\kappa_\nu \vec{F}_\nu + \int c\eta_\nu n^i d\Omega = \frac{\partial \vec{F}_\nu}{\partial t} + c^2\frac{\partial P_\nu}{\partial x^j} + 3H(t)\vec{F}_\nu$$
$$- H(t)\vec{F}_\nu - H(t)\nu\frac{\partial \vec{F}_\nu}{\partial \nu}. \quad (A13)$$

The emission from sources is isotropic, leading the integral to be zero, and ignoring the frequency dependence of flux we get

$$-c\kappa_\nu \vec{F}_\nu = \frac{\partial \vec{F}_\nu}{\partial t} + c^2\nabla P_\nu + 2H(t)\vec{F}_\nu. \quad (A14)$$

### A2 Frequency binning

Equations (A10) and (A14) are continuous in frequency. Consequently, these equations must be discretized before computation. The entire frequency range is divided into a series of discrete frequency bins. The radiative transfer equations are solved independently for each discretized frequency, with computational time scaling linearly with the number of frequency bins.

We start by defining the number density and flux as $N_i(F_i) = \int_{\nu_i}^{\nu_f} N_\nu(F_\nu)d\nu$, replacing the absorption coefficients with $\kappa_\nu = n_X\sigma_\nu$, where $\nu_i$ and $\nu_f$ are the lower and upper limits of the frequency bin, $n_X$ is the number density of the absorbing species, $\sigma$ is the cross-section for ionization. For a particular frequency bin after replacing the source terms and the absorption terms we get,

$$\frac{\partial N_i}{\partial t} + \nabla \vec{F}_i + 3H(t)N_i = -\sum_j^{\text{H\,I,He\,I,He\,II}} cn_j\sigma_{i,j}^N N_\nu + \dot{N}_i^* + N_i^{\text{rec}}, \quad (A15)$$

$$\frac{\partial \vec{F}_i}{\partial t} + c^2\nabla P_i + 2H(t)\vec{F}_i = -\sum_j^{\text{H\,I,He\,I,He\,II}} cn_j\sigma_{i,j}^E \vec{F}_i, \quad (A16)$$

where $\dot{N}_i^*$ is the rate of photon density from the stellar sources and $N_i^{\text{rec}}$ is rate of photon density from the recombination radiation from the gas. Assuming a frequency distribution for the radiative sources, which in our current setup is assumed to be that of a black-body, the number-weighted cross-sections $\sigma^N$ and energy-weighted cross-section $\sigma^E$ can be written as

$$\sigma_{i,j}^N = \frac{\int_{\nu_i}^{\nu_f}\sigma_{\nu j}I(\nu)/h\nu d\nu}{\int_{\nu_{i0}}^{\nu_{i1}}I(\nu)/h\nu d\nu}, \quad (A17)$$

$$\sigma_{i,j}^E = \frac{\int_{\nu_i}^{\nu_f}\sigma_{\nu j}I(\nu)d\nu}{\int_{\nu_{i0}}^{\nu_{i1}}I(\nu)d\nu}. \quad (A18)$$

### A3 M1 closure

Moment-taking of the equation can theoretically continue indefinitely, but must be truncated at some point. We stop at the first moment, which determines the second-order tensor found in equation (A14). This truncation leads to the introduction of the Eddington tensor. In the case of using the M1 closure relation, the Eddington tensor depends solely on the local radiation flux and intensity (Levermore 1984). The incorporation of this definition into equation (A14) transforms the equations into a hyperbolic system of conservation equations, the solutions to which are readily available. The general form of the Eddington tensor is as follows,

$$P_i = D_i \times N_i, \quad (A19)$$

$$D_i = \frac{1-\chi}{2}\vec{I} + \frac{3\chi-1}{2}n_i \otimes n_i. \quad (A20)$$

where $D_i$ is the Eddington tensor, $\chi$ is the Eddington factor, and $n_i$ is the unit vector aligned with the flux direction. The functional form for $\chi$ is defined as the M1 closure relation and is given by

$$\chi = \frac{3+4|f|^2}{5+2\sqrt{4-3|f|^2}}, \quad (A21)$$

where $f$ is the reduced flux given by $F/cN$. The efficacy of this method was examined in Wu, McQuinn & Eisenstein (2021), which concluded that the form of the Eddington tensor does not significantly influence how reionization progresses in cosmological RT simulations.

### A4 Thermochemistry equations

The interaction of species within the IGM with photons occurs in three ways: recombinations, photoionization, and collisional ionization. The net effect of these processes determines the time-dependent change of ionization states. The corresponding temperature-dependent coefficients, denoted by $\alpha$ and $\beta$ (taken from Hui & Gnedin 1997) can be found in Appendix A11. The variable $n_e$ represents the electron number density and depends on the species that are present in the simulation.

### A5 Hydrogen

As hydrogen has a single ionization state, only a single equation needs to be solved:

$$n_H\frac{dx_{\text{H\,II}}}{dt} = -\alpha_{\text{H\,II}}^A n_e n_{\text{H\,II}} + n_{\text{H\,I}}\left(\beta_{\text{H\,I}}n_e + \sum_{i=1}^M \sigma_{i\text{H\,I}}^N cN_i\right), \quad (A22)$$

where $M$ are the number of frequency bins, $x_{\text{H\,II}} = n_{\text{H\,II}}/n_H$ and $n_{\text{H\,II}} + n_{\text{H\,I}} = n_H$.

### A6 Helium

To incorporate helium into the radiative transfer scheme, the evolution of the ionization state of helium is solved concurrently with that of hydrogen. Helium has three different ionization states (He I, He II, and He III). Given that the total occupation fraction of these ionization states must equal one, we have a constraint and thus only need to solve for two out of the three ionization states. We choose to solve for He II and He III,

$$n_{\rm He}\frac{dx_{\rm He\,II}}{dt} = n_{\rm He\,I}\left(\beta_{\rm He\,I}n_e + \sum_{i=1}^{M}\sigma^N_{i\rm He\,I}cN_i\right) + n_{\rm He\,III}\alpha^A_{\rm He\,III}n_e$$
$$-n_{\rm He\,II}\left(\beta_{\rm He\,II}n_e + \alpha^A_{\rm He\,II}n_e + \sum_{i=1}^{M}\sigma^N_{i\rm He\,II}cN_i\right), \quad \text{(A23)}$$

$$n_{\rm He}\frac{dx_{\rm He\,III}}{dt} = n_{\rm He\,II}\left(\beta_{\rm He\,II}n_e + \sum_{i=1}^{M}\sigma^N_{i\rm He\,II}cN_i\right) - n_{\rm He\,III}\alpha^A_{\rm He\,III}n_e, \quad \text{(A24)}$$

where $x_{\rm He\,II} = n_{\rm He\,II}/n_{\rm He}$, $x_{\rm He\,III} = n_{\rm He\,III}/n_{\rm He}$, and $n_{\rm He\,I} + n_{\rm He\,II} + n_{\rm He\,III} = n_{\rm He}$.

### A7 Thermal energy density

Photon absorption and emission can both heat and cool the gas, making it necessary to solve the thermal energy consistently with the ionized fractions. This leads to,

$$\frac{dE}{dt} = \mathcal{H} - \mathcal{L} - 2HE,$$
$$= \sum_{i}^{M}\sum_{j}^{\rm H\,I,He\,I,He\,II} n_j cN_i(\epsilon_i\sigma^E_{ij} - \epsilon_j\sigma^N_{ij}) - \mathcal{L} - 2HE, \quad \text{(A25)}$$

where $\mathcal{H}$ is the photoheating rate, $\mathcal{L}$ is the radiative cooling rate, $\epsilon_i$ is the energy of the frequency bin and $\epsilon_j$ is the ionization energy of the species. The cooling term in the energy equation consists of the cooling rates due to collisional ionization, collisional excitations (Cen 1992), recombinations (Hui & Gnedin 1997), Bremsstrahlung (Osterbrock & Ferland 2006), dielectric recombination (Black 1981) and Compton cooling/heating (Haiman, Thoul & Loeb 1996).

### A8 Implementing multifrequency with helium in ATON

The successful implementation of the radiative transfer scheme involves solving a set of equations from (A10) to (A25). As previously discussed, ATON-HE accomplishes this using the M1 closure method. Within ATON-HE, the radiative process is segmented into two steps: photon transport and thermochemistry with photon injection. An operator splitting method is employed at each step for solving the differential equations. Furthermore, global time-step is determined using the Courant condition. The Courant condition ensures that photons do not traverse more than one volume element per time-step, thereby ensuring that $3c\Delta t/\Delta x < 1$. This condition relies on explicit integration where $m = n$ (see equation A30).

### A9 The photon transport step

This step focuses on the spatial propagation of photons. The equations describing the free-flowing photons are the homogeneous versions of equation (A10) and (A14):

$$\frac{\partial N_b}{\partial t} + \nabla \vec{F}_b = 0, \quad \text{(A26)}$$

$$\frac{\partial \vec{F}_b}{\partial t} + c^2\nabla P_b = 0, \quad \text{(A27)}$$

where subscript $b$ represents the frequency bin in consideration. Upon discretization of (A26) and (A27), we obtain

$$\frac{(N_b)^{n+1}_c - (N_b)^n_c}{\Delta t} + \frac{(F_b)^m_{c+1/2} - (F_b)^m_{c-1/2}}{\Delta x} = 0, \quad \text{(A28)}$$

$$\frac{(F_b)^{n+1}_c - (F_b)^n_c}{\Delta t} + c^2\frac{(P_b)^m_{c+1/2} - (P_b)^m_{c-1/2}}{\Delta x} = 0. \quad \text{(A29)}$$

The Global Lax-Friedrich (GLF) flux function (González, Audit & Huynh 2007) is employed to attain the value at the inter-cell $c + 1/2$ where $\mathscr{F} = (F_b, c^2 P_b)^T$ and $\mathscr{U} = (N_b, F_b)^T$. This function caters to isotropic sources and is expressed as

$$(\mathscr{F}_{\rm GLF})^m_{c+1/2} = \frac{\mathscr{F}^m_c + \mathscr{F}^m_{c+1}}{2} - \frac{c}{2}(\mathscr{U}^m_{c+1} - \mathscr{U}^m_c). \quad \text{(A30)}$$

In ATON-HE, the energy density and each component of the radiation flux are three-dimensional arrays, and thus the solution to equations (A28) and (A29) using an explicit scheme ($m = n$), substituting equation (A30),

$$N^{n+1}_{i,j,k} = N^n_{i,j,k} - \frac{\Delta t}{2\Delta x}\left((F_x)^n_{i+1,j,k} - (F_x)^n_{i-1,j,k}\right)$$
$$-\frac{\Delta t}{2\Delta z}\left((F_z)^n_{i,j,k+1} - (F_z)^n_{i,j,k-1}\right)$$
$$-\frac{\Delta t}{2\Delta y}\left((F_y)^n_{i,j+1,k} - (F_y)^n_{i,j-1,k}\right)$$
$$-\frac{c\Delta t}{2\Delta x}\left(2N^n_{i,j,k} - N^n_{i+1,j,k} - N^n_{i-1,j,k}\right)$$
$$-\frac{c\Delta t}{2\Delta y}\left(2N^n_{i,j,k} - N^n_{i,j+1,k} - N^n_{i,j-1,k}\right)$$
$$-\frac{c\Delta t}{2\Delta z}\left(2N^n_{i,j,k} - N^n_{i,j,k+1} - N^n_{i,j,k-1}\right), \quad \text{(A31)}$$

where $i$, $j$, $k$ refer to the array indices in the $x$, $y$, and $z$ directions, respectively, and $n + 1$ is the next time-step. This equation is solved for each frequency bin separately. The evolution of the radiation flux in the $x$ direction, using an explicit scheme

$$(F_x)^{n+1}_{i,j,k} = (F_x)^n_{i,j,k} - c^2\left(P^n_{i+1,j,k} - P^n_{i-1,j,k}\right)\frac{\Delta t}{2\Delta x}$$
$$-c^2\left(P^n_{i,j+1,k} - P^n_{i,j-1,k}\right)\frac{\Delta t}{2\Delta y}$$
$$-c^2\left(P^n_{i,j,k+1} - P^n_{i,j,k-1}\right)\frac{\Delta t}{2\Delta z}$$
$$-\left(2(F_x)^n_{i,j,k} - (F_x)^n_{i+1,j,k} - (F_x)^n_{i-1,j,k}\right)\frac{c\Delta t}{2\Delta x}$$
$$-\left(2(F_x)^n_{i,j,k} - (F_x)^n_{i,j+1,k} - (F_x)^n_{i,j-1,k}\right)\frac{c\Delta t}{2\Delta y}$$
$$-\left(2(F_x)^n_{i,j,k} - (F_x)^n_{i,j,k+1} - (F_x)^n_{i,j,k-1}\right)\frac{c\Delta t}{2\Delta z}, \quad \text{(A32)}$$

where $P$ is a three-dimensional array defined using equation (A19). This equation is solved for $y$, and $z$ components of the radiation flux for each frequency bin.

### A10 The thermochemistry step and photon injection

The alteration in the number of photons resulting from the interaction between photons and species in the IGM is described by

$$\frac{\partial N_i}{\partial t} = -\sum_{j}^{\rm He\,I,He\,I,He\,II} n_j c\sigma^N_{ij}N_i$$
$$+ \sum_{j}^{\rm H\,II,He\,II,He\,III}(\alpha^A_j - \alpha^B_j)n_j n_e + \dot{N}^*_i - 3H(t)N_i. \quad \text{(A33)}$$

While the change in radiation flux is given by

$$\frac{\partial \vec{F}_i}{\partial t} = -\sum_{j}^{\text{H\,{\sc i}, He\,{\sc i}, He\,{\sc ii}}} n_j c \sigma_{ij}^N \vec{F}_i - 2H(t)\vec{F}_i, \quad (A34)$$

where $n_e = n_{\text{H\,{\sc ii}}} + n_{\text{He\,{\sc ii}}} + 2n_{\text{He\,{\sc iii}}}$ and $\alpha^A$, $\alpha^B$ are the case A and case B recombination coefficients. Discretization of the number density, and solving it implicitly leads to

$$N_i^{p+1} = \frac{N_i^p + \Delta t \sum_j^{\text{H\,{\sc ii}, He\,{\sc ii}, He\,{\sc iii}}} \left(\alpha_j^a - \alpha_j^b\right) n_j^p n_e^p + \Delta t \dot{N}_i^{*n}}{1 + c\Delta t \sum_j^{\text{H\,{\sc i}, He\,{\sc i}, He\,{\sc ii}}} n_j^p \sigma_{ij}^N + 3H\Delta t}. \quad (A35)$$

The radiation equation is solved using a different time-step in a process called subcycling. The $p$ superscript refers to the subcycling time, while $n$ is the global time for the simulation. The $i$ subscript refers to a particular frequency bin. This equation is solved for each point in the three-dimensional array. Discretization of the radiation flux, and solving it implicitly leads to

$$\boldsymbol{F}_i^{p+1} = \frac{\boldsymbol{F}_i^p}{1 + \Delta t \sum_j^{\text{H\,{\sc i}, He\,{\sc i}, He\,{\sc ii}}} n_j^p c \sigma_{ij}^N + 2H\Delta t}. \quad (A36)$$

Lastly, we discretize the equations detailed in Section A4, and solve it implicitly. Starting with the evolution of the ionized fraction of ionized hydrogen, we get

$$x_{\text{H\,{\sc ii}}}^{p+1} = \frac{x_{\text{H\,{\sc ii}}}^p + \Delta t \left(\beta_{\text{fhi}}^p n_e^p + \sum_{i=1}^M \sigma_{i\text{fhi}}^N c N_i^{p+1}\right)}{1 + \Delta t \left(\alpha_{\text{H\,{\sc ii}}}^A\right)^p n_e^p + \Delta t \left(\beta_{\text{H\,{\sc i}}}^p n_e^p + \sum_{i=1}^M \sigma_{i\text{H\,{\sc i}}}^N c N_i^{p+1}\right)}. \quad (A37)$$

For the evolution of the He\,{\sc ii}, we will define the numerator as

$$\text{numerator} = x_{\text{He\,{\sc ii}}}^p + \Delta t \left(\beta_{\text{He\,{\sc i}}}^p n_e^p + \sum_{i=1}^M \sigma_{i\text{He\,{\sc i}}}^N c N_i^{p+1}\right)$$
$$- \Delta t x_{\text{He\,{\sc iii}}}^p \left(\left(\beta_{\text{He\,{\sc i}}}^p n_e^p + \sum_{i=1}^M \sigma_{i\text{He\,{\sc i}}}^N c N_i^{p+1}\right) - \left(\alpha_{\text{He\,{\sc iii}}}^A\right)^p n_e^p\right),$$

while the denominator is

$$\text{denominator} = 1 + \Delta t \left[\beta_{\text{He\,{\sc i}}}^p n_e^p + \sum_{i=1}^M \sigma_{i\text{He\,{\sc i}}}^N c N_i^{p+1}\right]$$
$$+ \Delta t \left[\beta_{\text{He\,{\sc ii}}}^p n_e^p + \left(\alpha_{\text{He\,{\sc ii}}}^A\right)^p n_e^p + \sum_{i=1}^M \sigma_{i\text{He\,{\sc ii}}}^N c N_i^{p+1}\right].$$

The updated value of the He\,{\sc ii} is $x_{\text{He\,{\sc ii}}} = $ numerator/denominator. Solving the evolution of He\,{\sc iii} implicitly we get,

$$x_{\text{He\,{\sc iii}}}^{p+1} = \frac{x_{\text{He\,{\sc iii}}}^p + \Delta t x_{\text{He\,{\sc ii}}}^{p+1} \left(\beta_{\text{He\,{\sc ii}}}^p n_e^p + \sum_{i=1}^M \sigma_{i\text{He\,{\sc ii}}}^N c N_i^{p+1}\right)}{1 + \Delta t \left(\alpha_{\text{He\,{\sc iii}}}^A\right)^p n_e^p}. \quad (A38)$$

Finally, we solve the temperature evolution using an implicit scheme on equation (A25) to get the numerator as

$$\text{numerator} = \frac{3}{2} k_B n_{\text{tot}}^p T^p - \Delta t \, \text{Cool}$$
$$+ \Delta t \sum_{i}^{M} \sum_{j}^{\text{H\,{\sc i}, He\,{\sc i}, He\,{\sc ii}}} n_j^{p+1} c N_i^{p+1} (\epsilon_i \sigma_{ij}^E - \epsilon_j \sigma_{ij}^N), \quad (A39)$$

where $n_{\text{tot}} = n_{\text{H}} + n_{\text{He}} + n_e$, $n_{\text{H}}$ is the number density of hydrogen, $n_{\text{He}}$ is the number density of helium, and $n_e$ is the number density of electrons. The 'Cool' is the total cooling rate defined as

$$\text{Cool} = \zeta_{\text{H\,{\sc i}}} n_e n_{\text{H\,{\sc i}}} + \zeta_{\text{He\,{\sc i}}} n_e n_{\text{He\,{\sc i}}} + \zeta_{\text{He\,{\sc ii}}} n_e n_{\text{He\,{\sc ii}}} + \eta_{\text{H\,{\sc ii}}}^A n_e n_{\text{H\,{\sc ii}}}$$
$$+ \eta_{\text{He\,{\sc ii}}}^A n_e n_{\text{He\,{\sc ii}}} + \eta_{\text{He\,{\sc iii}}}^A n_e n_{\text{He\,{\sc iii}}} + \psi_{\text{H\,{\sc i}}} n_e n_{\text{H\,{\sc i}}} + \psi_{\text{He\,{\sc ii}}} n_e n_{\text{He\,{\sc ii}}}$$
$$+ \theta n_e (n_{\text{H\,{\sc ii}}} + n_{\text{He\,{\sc ii}}} + 4n_{\text{He\,{\sc iii}}}) + \omega_{\text{He\,{\sc ii}}} n_e n_{\text{He\,{\sc ii}}} + \varpi n_e. \quad (A40)$$

The temperature coefficients are defined in Section A11. The denominator is

$$\text{denominator} = \left(\frac{3}{2} k_B n_{\text{tot}}^{p+1}\right) (1 + 2H\Delta t) \quad (A41)$$

to get $T^{p+1} = $ numerator/denominator.

### A11 Temperature dependent coefficients

We use the following values of rate coefficients for case A and case B recombination (Hui & Gnedin 1997):

$$\alpha_{\text{H\,{\sc ii}}}^A = 1.269 \times 10^{-13} \text{ cm}^3 \text{ s}^{-1} \frac{\lambda_{\text{H\,{\sc i}}}^{1.503}}{[1 + (\lambda_{\text{H\,{\sc i}}}/0.522)^{0.47}]^{1.923}}, \quad (A42)$$

$$\alpha_{\text{He\,{\sc ii}}}^A = 3 \times 10^{-14} \text{ cm}^3 \text{ s}^{-1} \lambda_{\text{He\,{\sc i}}}^{0.654}, \quad (A43)$$

$$\alpha_{\text{He\,{\sc iii}}}^A = 2.538 \times 10^{-13} \text{ cm}^3 \text{ s}^{-1} \frac{\lambda_{\text{He\,{\sc ii}}}^{1.503}}{[1 + (\lambda_{\text{He\,{\sc ii}}}/0.522)^{0.47}]^{1.923}}, \quad (A44)$$

$$\alpha_{\text{H\,{\sc ii}}}^B = 2.753 \times 10^{-14} \text{ cm}^3 \text{ s}^{-1} \frac{\lambda_{\text{H\,{\sc i}}}^{1.5}}{[1 + (\lambda_{\text{H\,{\sc i}}}/2.74)^{0.407}]^{2.242}}, \quad (A45)$$

$$\alpha_{\text{He\,{\sc ii}}}^B = 1.26 \times 10^{-14} \text{ cm}^3 \text{ s}^{-1} \lambda_{\text{He\,{\sc i}}}^{0.75}, \quad (A46)$$

$$\alpha_{\text{He\,{\sc iii}}}^B = 5.506 \times 10^{-14} \text{ cm}^3 \text{ s}^{-1} \frac{\lambda_{\text{He\,{\sc ii}}}^{1.5}}{[1 + (\lambda_{\text{He\,{\sc ii}}}/2.74)^{0.407}]^{2.242}}. \quad (A47)$$

Here, we have $\lambda_j = 2T_j/T$, where $j$ indexes the three species H\,{\sc i}, He\,{\sc i}, and He\,{\sc ii}. The values for $T_j$ are $T_{\text{H\,{\sc i}}} = 157807$ K, $T_{\text{He\,{\sc i}}} = 285335$ K, and $T_{\text{He\,{\sc ii}}} = 631515$ K. We use the following values for the collisional ionization rate coefficients (Cen 1992) with temperature in Kelvin

$$\beta_{\text{H\,{\sc i}}} = 5.85 \times 10^{-11} \text{ cm}^3 \text{ s}^{-1} \sqrt{T} \left(1 + \sqrt{\frac{T}{10^5}}\right)^{-1} e^{-157809.1/T}, \quad (A48)$$

$$\beta_{\text{He\,{\sc i}}} = 2.38 \times 10^{-11} \text{ cm}^3 \text{ s}^{-1} \sqrt{T} \left(1 + \sqrt{\frac{T}{10^5}}\right)^{-1} e^{-285335.4/T}. \quad (A49)$$

We use the following values for the recombination cooling rate coefficients (Hui & Gnedin 1997), with temperatures in Kelvin, and the Boltzmann constant in erg K$^{-1}$,

$$\eta_{\text{H\,{\sc ii}}}^A = 1.778 \times 10^{-29} \text{ erg cm}^3 \text{ s}^{-1} \text{ K}^{-1} \frac{\lambda_{\text{H\,{\sc i}}}^{1.965} T}{\left[1.0 + \frac{\lambda_{\text{H\,{\sc i}}}}{0.541}^{0.502}\right]^{2.697}}, \quad (A50)$$

$$\eta_{\text{He\,{\sc ii}}}^A = 3.0 \times 10^{-14} \text{ cm}^3 \text{ s}^{-1} \lambda_{\text{He\,{\sc i}}}^{0.654} k_B T, \quad (A51)$$

$$\eta_{\text{He\,{\sc iii}}}^A = 8 \times 1.778 \times 10^{-29} \text{ erg cm}^3 \text{ s}^{-1} \text{ K}^{-1} \frac{\lambda_{\text{He\,{\sc ii}}}^{1.965} T}{\left[1.0 + \frac{\lambda_{\text{He\,{\sc ii}}}}{0.541}^{0.502}\right]^{2.697}}, \quad (A52)$$

$$\eta_{\text{H\,{\sc ii}}}^B = 3.435 \times 10^{-30} \text{ erg cm}^3 \text{ s}^{-1} \text{ K}^{-1} \frac{\lambda_{\text{H\,{\sc i}}}^{1.970} T}{\left[1.0 + \frac{\lambda_{\text{H\,{\sc i}}}}{2.250}^{0.376}\right]^{3.720}}, \quad (A53)$$

$$\eta_{\text{He\,{\sc ii}}}^B = 1.26 \times 10^{-14} \text{ cm}^3 \text{ s}^{-1} \lambda_{\text{He\,{\sc i}}}^{0.750} k_B T, \quad (A54)$$

$$\eta_{\text{He\,{\sc iii}}}^B = 8 \times 3.435 \times 10^{-30} \text{ erg cm}^3 \text{ s}^{-1} \text{ K}^{-1} \frac{\lambda_{\text{He\,{\sc ii}}}^{1.970} T}{\left[1.0 + \frac{\lambda_{\text{He\,{\sc ii}}}}{2.250}^{0.376}\right]^{3.720}}. \quad (A55)$$

We use the following values for the cooling rate coefficients (with units erg cm$^3$ s$^{-1}$) for collisional ionization and collisional excitations (Cen 1992),

$$\zeta_{\text{H\,I}} = 1.27 \times 10^{-21} \sqrt{T} \left(1 + \sqrt{\frac{T}{10^5}}\right)^{-1} e^{-157809.1/T}, \quad \text{(A56)}$$

$$\zeta_{\text{He\,I}} = 9.38 \times 10^{-22} \sqrt{T} \left(1 + \sqrt{\frac{T}{10^5}}\right)^{-1} e^{-285335.4/T}, \quad \text{(A57)}$$

$$\zeta_{\text{He\,II}} = 4.95 \times 10^{-22} \sqrt{T} \left(1 + \sqrt{\frac{T}{10^5}}\right)^{-1} e^{-631515/T}, \quad \text{(A58)}$$

$$\psi_{\text{H\,I}} = 7.5 \times 10^{-19} \left(1 + \sqrt{\frac{T}{10^5}}\right)^{-1} e^{-118348/T}, \quad \text{(A59)}$$

$$\psi_{\text{He\,II}} = 5.54 \times 10^{-17} T^{-0.397} \left(1 + \sqrt{\frac{T}{10^5}}\right)^{-1} e^{-473638/T}. \quad \text{(A60)}$$

The bremsstrahlung cooling rate coefficients are given by (Osterbrock & Ferland 2006),

$$\theta_{\text{H\,II}} = 1.42 \times 10^{-27} \text{ erg cm}^3 \text{ s}^{-1} \sqrt{T}, \quad \text{(A61)}$$

$$\theta_{\text{He\,II}} = 1.42 \times 10^{-27} \text{ erg cm}^3 \text{ s}^{-1} \sqrt{T}, \quad \text{(A62)}$$

$$\theta_{\text{He\,III}} = 4 \times 1.42 \times 10^{-27} \text{ erg cm}^3 \text{ s}^{-1} \sqrt{T}. \quad \text{(A63)}$$

The Compton cooling rate coefficient, with $a$ being the scale factor is given by Haiman et al. (1996),

$$\varpi(T, a) = 1.017 \times 10^{-37} \text{ erg s}^{-1} \left(\frac{2.727}{a}\right)^4 \left(T - \frac{2.727}{a}\right). \quad \text{(A64)}$$

The dielectric recombination cooling rate coefficient (with units erg cm$^3$ s$^{-1}$) is given by (Black 1981)

$$\omega_{\text{He\,II}} = 1.24 \times 10^{-13} T^{-1.5} e^{-470000/T} \times \left(1 + 0.3 e^{-94000/T}\right). \quad \text{(A65)}$$

## APPENDIX B: CODE TEST

We tested ATON-HE using the well-established Strömgren sphere problem. A point source is placed in a uniform density distribution of hydrogen and helium. For a hydrogen-only medium it is possible to compare the result with the well-known analytical solution, but for a medium consisting of hydrogen and helium, an analytical solution does not exist. We therefore compare our computation with the results of a similar calculation by Friedrich et al. (2012) using C$^2$-RAY code, a non-equilibrium photoionization code that utilizes short-characteristic ray tracing for radiative transfer.

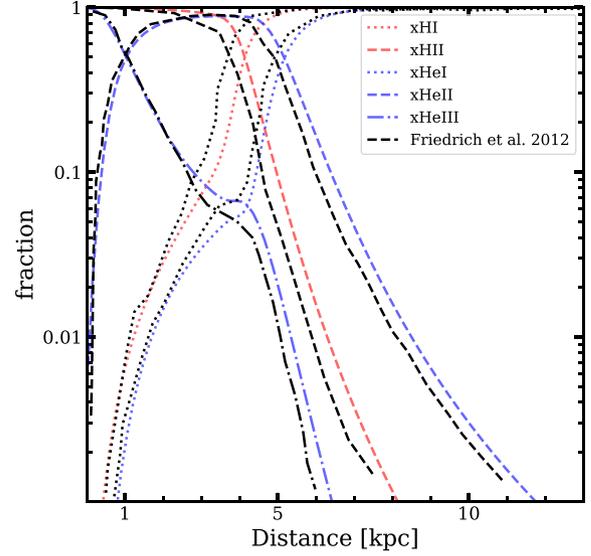

**Figure B2.** A comparison of the ionized fractions in our Strömgren sphere test run with results from Friedrich et al. (2012) at $t = 100$ Myr.

The Strömgren sphere simulations were established with the ionizing source situated at the centre of the simulation box, surrounded by a uniformly dense gas at a constant temperature. We adopted the same parameters as Friedrich et al. (2012), setting the hydrogen density of hydrogen in the medium to $n_{\text{H}} = 10^{-3}$ cm$^{-3}$, the number density of helium is 25 per cent by mass, and the initial gas temperature to $T_{\text{ini}} = 10^4$ K. The hydrogen ionizing photons emitted is set to $\dot{N}_\gamma = 5 \times 10^{48}$ s$^{-1}$. We choose a spatial resolution of $\Delta r = 150$ pc. Our simulation box spans 38.4 kpc with a grid size of 256. To avoid impractically long computation times due to the Courant condition, we reduced the speed of light to 0.01$c$. We used a black-body spectrum with $T = 10^5$ K, with nine frequency bins. The average photon energy within each bin is 18.85, 27.17, 33.08, 40.24, 48.92, 60.09, 74.99, 93.48, and 118.14 eV. All parameter, except the speed of light and the number of frequency bins are identical to Friedrich et al. (2012).

In Fig. B1, we present slices (thickness is 150 pc) through the centre of the simulation box, showing the distribution of the ionized fractions at $t = 100$ Myr. As expected, the distribution of the ionized fraction is spherically symmetric around the centre.

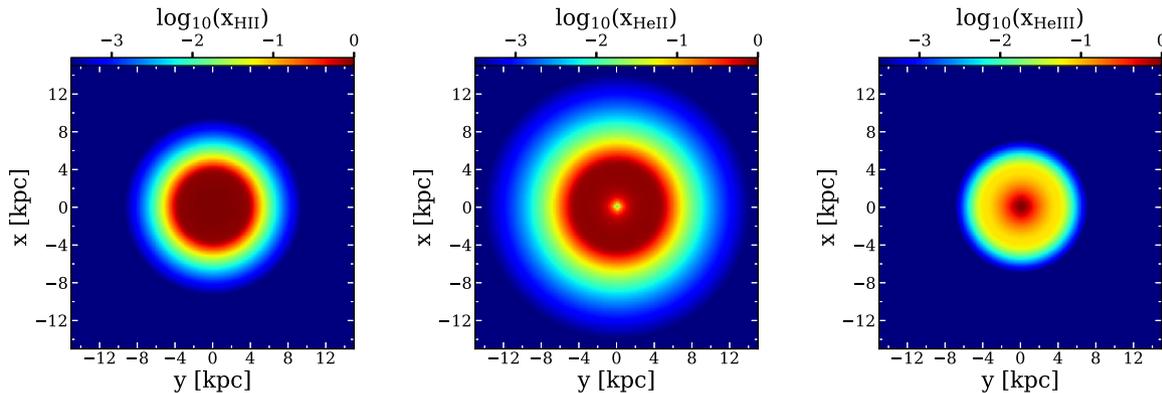

**Figure B1.** A two-dimensional slice through the centre of our Strömgren sphere simulation at $t = 100$ Myr, showing $x_{\text{H\,II}}$ (left panel), $x_{\text{He\,II}}$ (middle panel), and $x_{\text{He\,III}}$ (right panel). The box size is 38.4 kpc with a resolution of 150 pc. See Appendix B for more details.

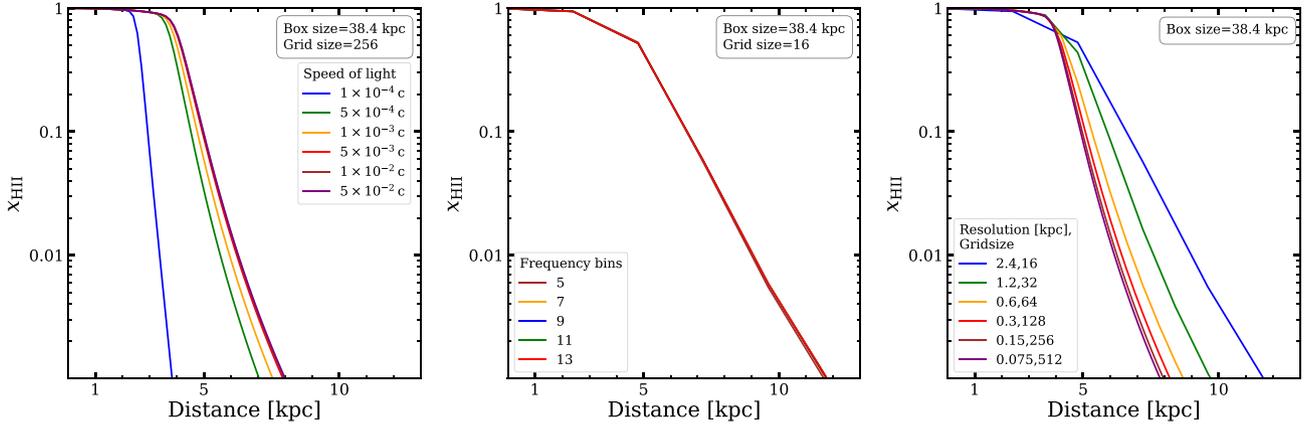

**Figure B3.** Effect of reduced speed of light (left panel), number of frequency bins (middle panel), and spatial resolution (right panel) on the ionized hydrogen fraction in the Strömgren test. The box size is 38.4 kpc in all panels. In the left panel, we use a uniform $256^3$ grid. In the central panel, the grid is $16^3$. The speed of light for the centre and right panels is $1 \times 10^{-2}$ c.

The ionized hydrogen fraction decreases monotonically as we move further away from the centre. Similar behaviour is observed for fraction of doubly ionized helium, although the extent of ionization differs due to the slightly higher recombination rates of He III and the relatively low flux at the relevant energies. For He II, we observe that the region with the highest value of the ionized fraction is not at the centre. This is because of the He III ionization front that starts to develop near the centre of the box. This behaviour of the ionization fractions is seen more clearly in Fig. B2, where we plot the one-dimensional distribution of the ionized fractions from Fig. B1, also at $t = 100$ Myr. Also shown in this figures are the results from Friedrich et al. (2012) for comparison. The results from the two codes compare well. The residual small differences can be attributed to the differences between the M1 closure technique of ATON-HE and the ray tracing used in C$^2$-RAY, as detailed in Iliev et al. (2009), and the implementation of the on-the-spot approximation by Friedrich et al. (2012).

We did a series of convergence tests to examine the effects of the speed of light, number of bins, and spatial resolution, on the above results. Fig. B3 shows the results of these tests. We see that our chosen values of these three parameters give converged results.

# APPENDIX C: ACCURACY REQUIREMENTS ON FLUX CALIBRATION

Our simulations are calibrated to mean Ly α transmission measurements. However, even in the calibrated runs, the agreement between the simulated Ly α transmission with its observed value is not perfect. This is often because manually finding the required emissivity evolution is challenging. But inaccuracies can also arise because the simulation boxes are not always available at the exact redshift values corresponding to the observations. One can therefore ask what the acceptable level of agreement should be between the simulated and observed values of the mean Ly α transmission. We investigate this in Fig. C1. Panel (a) of these figures show the simulated mean Ly α flux from 10 runs. The orange curve is our chosen calibration, while the blue curves represent the range of the other calibrations. All simulations are consistent with the data given the observational uncertainties. But, as panel (b) shows, the mean flux in the runs are different up to about 10 per cent. Panels (c) and (d) show the CDFs of the effective Ly α opacity for the different calibrations at the redshifts where the differences in mean flux are largest. The measured CDFs are also shown for context. We see that the 10 per cent differences in the mean flux do not magnify to large differences in the CDFs.

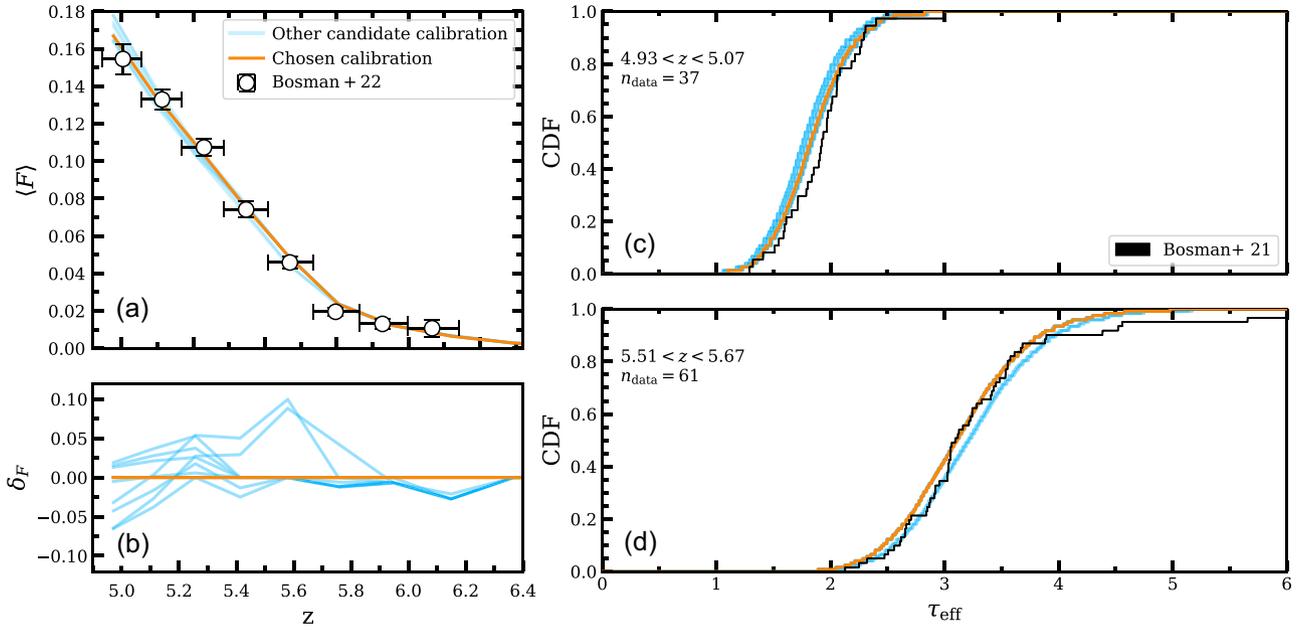

**Figure C1.** Panel (a) illustrates how the mean flux varies for consistent calibrations when the total volume emissivity is adjusted. The orange line is for our fiducial calibration, while the blue curves show possible alternative calibrations. Adjustments to the emissivity were made to ensure that the simulation results agree closely with the observed mean flux, as indicated by the black markers, based on (Bosman et al. 2022). Panel (b) shows the differences in the mean flux $\delta_F$ of the fiducial calibrated simulation (orange) and the rest of the simulations (the blue curves in Panel a). At the relevant redshifts, the largest differences between the different calibrations is around 10 per cent. Panels (c) and (d) show the distribution of CDFs of the effective optical depth at the two redshifts where the differences in the mean flux are largest. The orange curves show the median CDF for the fiducial calibration, while the blue curves show the range of median CDFs for the alternative calibrations. A variation of approximately 10 per cent in mean flux values corresponds to similar differences in the CDFs.

## APPENDIX D: THERMAL EVOLUTION CONVERGENCE

Fig. D1 illustrates the variation of temperature at mean density in an ATON-HE simulation with box size 20 cMpc/$h$ and a $512^3$ grid. The source spectrum is black-body with temperature $4 \times 10^4$ K. Different curves show the simulation result when all that is changed is the number of energy bins (See Section 2.2.2 for a detailed description for frequency binning). We observe that as the number of frequency bins increases, the temperature suddenly rises by approximately 20 per cent upon the addition of the fourth bin, after which it stabilizes around that value. This is because with just three bins, the part of the source spectrum with energies greater than the He II ionizing potential is not sampled well enough. With number of bins greater than four, however, we see a convergence of about 5 per cent in the temperature computation. This motivates the use of four photon energy bins used in this paper.

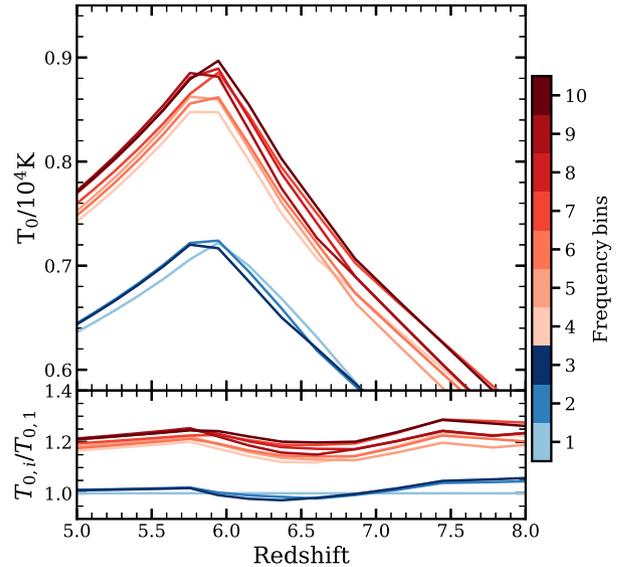

**Figure D1.** Top panel shows the evolution of the gas temperature at mean density for test simulations with a box size of 20 cMpc $h^{-1}$ and a $512^3$ uniform grid. The number of frequency bins range from one to ten. The bottom panel displays the ratio with the monofrequency result.

This paper has been typeset from a TEX/LATEX file prepared by the author.